\documentclass[10pt,a4paper,onecolumn]{article}
\usepackage[T1]{fontenc}
\usepackage{amssymb}
\usepackage[utf8]{inputenc}
\usepackage{setspace}
\usepackage[font=footnotesize]{caption}
\usepackage[biblabel]{cite}
\usepackage{graphicx}
\usepackage{float}
\usepackage{lineno}
\usepackage[dvipsnames]{xcolor}
\usepackage{placeins}
\usepackage{amsmath}
\usepackage{multicol}
\setpagewiselinenumbers
\let\oldbibliography\thebibliography
\renewcommand{\thebibliography}[1]{%
  \oldbibliography{#1}%
  \setlength{\itemsep}{0pt}%
}
\graphicspath{{Figs/}}

\begin{document}
\pagenumbering{arabic}

\begin{center}
{\Large{\bf {\Large{\bf Van der Waals Materials for Applications in Nanophotonics}}}}
\vskip1.0\baselineskip{Panaiot G. Zotev$^{a*}$, Yue Wang$^b$, Daniel Andres-Penares$^c$, Toby Severs Millard$^a$, Sam Randerson$^a$, Xuerong Hu$^a$, Luca Sortino$^d$, Charalambos Louca$^a$, Mauro Brotons-Gisbert$^c$, Tahiyat Huq$^e$, Stefano Vezzoli$^e$, Riccardo Sapienza$^e$, Thomas F. Krauss$^b$, Brian D. Gerardot$^c$, Alexander I. Tartakovskii$^{a**}$}
\vskip0.5\baselineskip\footnotesize{\em$^a$Department of Physics and Astronomy, University of Sheffield, Sheffield, S3 7RH, UK\\$^b$School of Physics, Engineering and Technology, University of York, York, YO10 5DD, UK\\$^c$School of Engineering and Physical Sciences, Heriot-Watt University, Edinburgh, EH14 4AS, UK\\$^d$Chair in Hybrid Nanosystems, Nanoinstitute Munich, Faculty of Physics, Ludwig-Maximilians-Universit\"{a}t, 80539, Munich, Germany\\$^e$The Blackett Laboratory, Department of Physics, Imperial College London, London, SW7 2BW, UK\\}
$^{*}$p.zotev@sheffield.ac.uk \quad $^{**}$a.tartakovskii@sheffield.ac.uk

\end{center}
\vskip1.0\baselineskip

\begin{multicols}{2}

\textbf{Numerous optical phenomena and applications have been enabled by nanophotonic structures. Their current fabrication from high refractive index dielectrics, such as silicon or gallium phosphide, pose restricting fabrication challenges, while metals, relying on plasmons and thus exhibiting high ohmic losses, limit the achievable applications. Here, we present an emerging class of layered so-called van der Waals (vdW) crystals as a viable nanophotonics platform. We extract the dielectric response of 11 mechanically exfoliated thin-film (20-200 nm) van der Waals crystals, revealing high refractive indices up to n = 5, pronounced birefringence up to $\Delta$n = 3, sharp absorption resonances, and a range of transparency windows from ultraviolet to near-infrared. We then fabricate nanoantennas on SiO$_2$ and gold utilizing the compatibility of vdW thin films with a variety of substrates. We observe pronounced Mie resonances due to the high refractive index contrast on SiO$_2$ leading to a strong exciton-photon coupling regime as well as largely unexplored high-quality-factor, hybrid Mie-plasmon modes on gold. We demonstrate further vdW-material-specific degrees of freedom in fabrication by realizing nanoantennas from stacked twisted crystalline thin-films, enabling control of nonlinear optical properties, and post-fabrication nanostructure transfer, important for nano-optics with sensitive materials.}\\

Nanophotonics in the 21st century has aimed to efficiently confine and manipulate light far below its wavelength using nanoscale structures. The motivation for fabricating these nanostructures lies in the applications enabled by their resonances in the visible and near-infrared portions of the electromagnetic spectrum including waveguiding in optical circuits \cite{Jamois2003,Rigal2018}, Purcell enhancement of emission \cite{Lodahl2015,Cambiasso2017,Liu2018b,Sortino2019}, low-threshold lasing \cite{Wu2015}, higher harmonic generation enhancement \cite{Cambiasso2017,Grinblat2017}, strong light-matter coupling \cite{Chikkaraddy2016,Baranov2018} and optical trapping \cite{Wang2011,Xu2018,Conteduca2021}.

For more than a decade, this field of research largely focused on nanostructures fabricated from noble metals, which host plasmonic modes in nanoparticles \cite{Chikkaraddy2016}, nanoantennas \cite{Akselrod2014}, waveguides \cite{Blauth2018} and nanocubes \cite{Luo2018}. While such metallic structures achieve vanishingly small mode volumes \cite{Chikkaraddy2016} thereby increasing the light-matter interaction, they suffer from high ohmic losses \cite{Khurgin2015}, cannot maintain magnetic resonances \cite{Bakker2015a} and induce sample heating which may be detrimental for biological applications \cite{Xu2018}. An alternative group of materials which provide solutions to these problems are dielectrics \cite{Baranov2017} such as silicon (Si) \cite{Lodahl2015,Bakker2015a,Xu2018}, germanium \cite{Grinblat2017}, gallium arsenide \cite{Lodahl2015,Liu2018b}, and gallium phosphide \cite{Cambiasso2017,Sortino2019}.

Despite the many advances offered by traditional plasmonic and dielectric nanostructures, limitations in refractive index, fabrication difficulty and versatility remain, yet can be resolved by the use of thin-film van der Waals materials. Similar to other dielectrics, layered materials do not suffer from ohmic losses, their nanostructures can maintain magnetic resonances \cite{Verre2019} and do not induce detrimental sample heating. Due to a general rule concerning the inverse relation of the refractive index of dielectrics and their bandgap ($n \approx E_{g}^{-1/4}$) \cite{Moss1985}, limitations in the number of materials available for fabricating nanostructures without absorption in the visible range remain. Van der Waals materials can provide a solution as they offer larger refractive indices (n>4) in this portion of the spectrum \cite{Verre2019,Munkhbat2022a}, a range of transparency windows well into the ultraviolet \cite{Rah2019} and numerous advantages due to their van der Waals adhesive nature to a variety of substrates without the necessity of lattice matching, a well known constraint of traditional dielectric nanophotonics \cite{Liu2019}. Emerging from this, hybrid plasmonic-dielectric nanoresonators, yielding low optical losses and large photonic enhancement factors \cite{Yang2017}, may become relatively simple to realize by transfer of a layered material onto a metallic surface followed by patterning of a nanophotonic structure. Another approach, achievable due to the weak van der Waals adhesion, is the fabrication of nanophotonic structures from stacked layers of the same or different materials with mutually twisted crystal axes which have applications in nonlinear optics \cite{Yao2021}. This builds on previous work from the widely studied van der Waals heterostrocture realizations \cite{Geim2013}. Additionally, post-fabrication techniques for designing bespoke nano-devices, such as repositioning via an atomic force microscope (AFM) tip \cite{Zotev2022} is readily available to vdW photonic structures, while not applicable to most structures made from traditional high refractive index dielectrics.

Recent works have demonstrated the fabrication of photonic crystal cavities, ring resonators, circular Bragg gratings and waveguides from hexagonal boron nitride (hBN) \cite{Caldwell2019}. Nanoantennas \cite{Verre2019,Busschaert2020,Zotev2022}, photonic crystals \cite{Zhang2020}, microring resonators and rib waveguides \cite{Munkhbat2022} as well as metasurfaces \cite{Munkhbat2022,Qin2022,Weber2022} have been fabricated from only two transition metal dichalcogenides (TMDs), namely WS$_2$ and MoS$_2$. These structures have been used to demonstrate a range of applications including strong light-matter coupling \cite{Verre2019,Zhang2020,Zhang2020a,Qin2022,Weber2022}, waveguiding \cite{Ermolaev2021,Munkhbat2022}, Purcell enhancement of emission \cite{Zotev2022} and higher harmonic generation enhancement \cite{Busschaert2020,Popkova2022,Zotev2022,Xu2022}. However, realizations of nanophotonic architectures using further vdW materials remain elusive despite numerical studies exploring exciting future applications in BICs \cite{Muhammad2021} as well as entire optical circuits \cite{Ling2021}.

In order to inspire and facilitate the use of a larger range of vdW materials for the fabrication of nanophotonic structures, in this work, we study the optical properties of a variety of layered materials and characterize their utility in different applications. We extract the dielectric response of each material via micro-ellipsometry, yielding large refractive indices (n>4) in the visible with a range of transparency windows from the near-infrared to the ultraviolet. We observe transparency in the out-of-plane orientation as well as large birefringence values ($\Delta n \approx$ 3) for a number of layered materials. 

We pattern single (monomer) and double (dimer) nanoantenna resonators into a range of vdW materials. Studying the resulting geometry of the nanoantennas provides insight into the etching speed of crystal axes in different materials. We also fabricate nanoantenna structures into twisted stacks of thin-film WS$_2$ crystals demonstrating an ability to fabricate nanophotonic homostructures in which optical properties, such as second harmonic generation (SHG), can be controlled via the twist angle. 

The fabricated structures yield strong photonic resonances formed due to a large refractive index mismatch achievable as a result of the inherent ease of fabricating vdW structures on a SiO$_2$ substrate. This leads to the observation of strong light-matter coupling at room temperature exhibiting large Rabi splittings of > 100 meV in single nanoantennas of different TMDs.

We subsequently demonstrate the versatility of vdW material nanofabrication by etching WS$_2$ monomer nanoantennas directly onto a gold substrate, realizing hybrid Mie-plasmonic resonances with high quality factors \cite{Yang2017}. We also demonstrate the ability to recreate these nanostructures by employing a post-fabrication pick up and transfer technique, which we name "transferable photonics", useful for coupling nanoresonators to sensitive materials, such as TMD monolayers or biological systems.

As nonlinear light applications employing layered materials have drawn a significant interest in the past \cite{Busschaert2020,Popkova2022,Zotev2022,Xu2022}, we further characterize the third harmonic generation (THG) susceptibilities of several thin-film TMDs and demonstrate the application of THG enhancement in WSe$_2$ nanoantennas.\\

\large
\textbf{Results}\\
\normalsize

\textbf{Linear optical properties of van der Waals materials} As a first step in characterizing vdW crystals for nanophotonics fabrication, we extract the refractive index and extinction coefficient of a number popular materials including insulating hBN, TMDs, III-VI and magnetic materials. We begin by mechanically exfoliating multilayer crystals of 11 different vdW materials onto a 97 nm SiO$_2$ on silicon substrate. We characterize the thickness of each crystal via AFM and measure the complex reflectance ratio of each sample via spectroscopic micro-ellipsometry, recording the amplitude ($\Psi$) and phase ($\Delta$) at three different angles of incidence (see supplementary Note 1), schematically shown in Figure \ref{F1}(a). Each set of data is fitted with an appropriate multilayer model yielding the real ($n$) and imaginary ($\kappa$) components of the complex refractive index shown in Figure \ref{F1}(b)-(l) for each material including many members of the transition metal dichalcogenide family (WS$_2$, WSe$_2$, MoS$_2$, MoSe$_2$, ZrSe$_2$, HfSe$_2$), which are faintly shaded in red; two III-VI materials (GaS, In$_2$Se$_3$), which are shaded in blue; two magnetic layered materials (MnPSe$_3$, NiPS$_3$), which are shaded in green; and the insulating hBN, which is shaded in yellow.

We use an analytical model appropriate to each material based on two considerations: the presence or absence of an absorption bandgap within the experimentally achievable wavelength range (360 - 1000 nm) and any anisotropy in the dielectric response due to crystallographic asymmetry. Therefore, we divided the 11 materials into three categories (see supplementary Note 1) which are: semiconductors with a large bandgap (GaS, hBN), TMDs (WS$_2$, WSe$_2$, MoS$_2$, MoSe$_2$, ZrSe$_2$, HfSe$_2$) and isotropic materials (In$_2$Se$_3$, MnPSe$_3$, NiPS$_3$). We observe very strong absorption resonances, leading to high refractive indices, for well studied TMDs such as WS$_2$, WSe$_2$, MoS$_2$ and MoSe$_2$ which has been confirmed by previous reports in monolayers \cite{Li2014a}. We also see prominent absorption resonances for ZrSe$_2$, HfSe$_2$, MnPSe$_3$ and NiPS$_3$ suggesting there may be a similar response in these materials. An interesting note here is that due to the strong binding energy of excitons in these materials, they are confined to individual layers and therefore, absorption processes can be ascribed to an in-plane regime while the out-of-plane regime remains transparent. This suggests that these materials are transparent for light propagating in-plane with an out-of-plane electric field. This anisotropy in the dielectric response results in especially high values of birefringence ($\Delta n \approx$ 3.0) observed for MoS$_2$ and ZrSe$_2$ near their lowest energy excitonic resonances in near-infrared wavelengths (see Supplementary Note 2). 

The extracted refractive indices and extinction coefficients for GaS, hBN, MoS$_2$, MoSe$_2$, WS$_2$, WSe$_2$ and In$_2$Se$_3$ are in good agreement with previously reported values \cite{Kato2011,Rah2019,Segura2018,Ermolaev2021,Munkhbat2022a,Cho2019}, corroborating our approach. The extinction coefficients we extract for HfSe$_2$, ZrSe$_2$ and NiPS$_3$, which to the best of our knowledge have not been previously reported, are consistent with DFT calculations of absorption in NiPS$_3$ \cite{Kim2018b}, HfSe$_2$ and ZrSe$_2$ \cite{Jaiswal2018}, as well as experimentally measured absorption in NiPS$_3$ \cite{Lane2020}.\\

\begin{figure}[ht!]
	\centering
  \includegraphics[width=\linewidth]{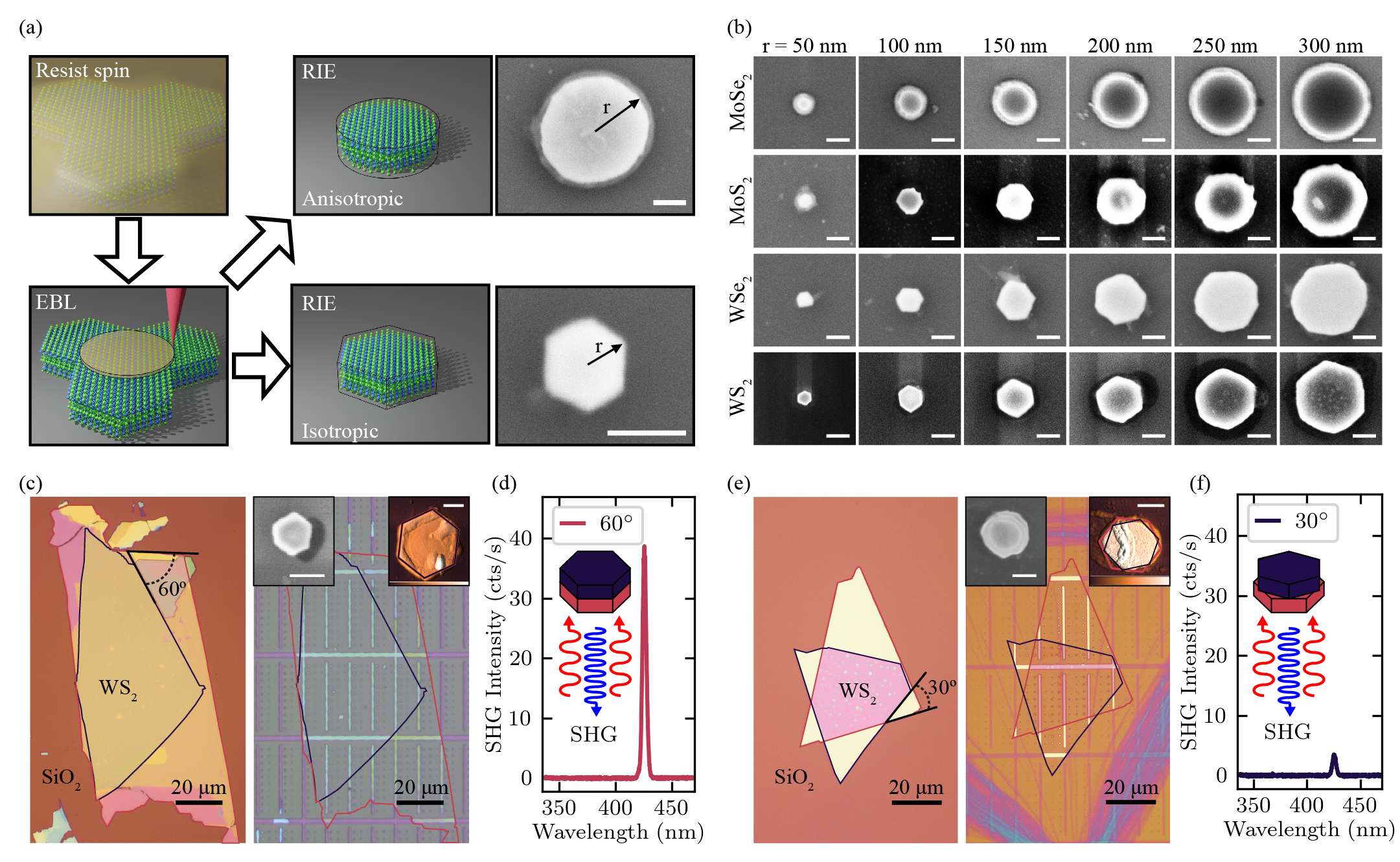}
  \caption{\textbf{Fabrication of vdW material nanopillars.}
	\textbf{(a)} Fabrication steps and their order following the black outline arrows. The first step includes spinning of resist onto a vdW crystal. The second step is patterning and development after electron beam lithography into a circular geometry. The final step is reactive ion etching using either an anisotropic or isotropic etching recipe. The height (h) of the resulting nanostructure is defined by the thickness of the original thin-film. The radius (r) is defined as the distance from the center of the nanostructure to an outside vertex for the hexagonal geometry.
	\textbf{(b)} Scanning electron microscopy (SEM) images of fabricated nanopillars in MoSe$_2$, MoS$_2$, WSe$_2$, WS$_2$ using the same isotropic etching recipe. WS$_2$ nanopillars maintain a hexagonal shape for all sizes while WSe$_2$ and MoS$_2$ exhibit circular geometries for a radius above 250 nm and 150 nm respectively. Isotropic etching of MoSe$_2$ nanopillars maintain a circular shape for all recorded sizes.
	\textbf{(c)} Optical images of a WS$_2$ homostructure achieved by the stacking of two thin-film crystals (thickness = 50 nm) with a 60$^\circ$ twist angle before (left panel) and after (right panel) fabrication of single nanopillars. The steps include the transfer of the crystals to form the homostructure on a SiO$_2$ substrate, EBL and RIE. Left inset: SEM image of a fabricated 60$^\circ$ twisted nanopillar. Right inset: AFM scan of a nanopillar in the 60$^\circ$ homostructure with hexagonal outlines to show the orientation of the top (black) and bottom (red) crystal. 
	\textbf{(d)} SHG spectrum collected from a 60$^\circ$ twisted nanopillar (r = 290 nm, h = 100 nm) under 850 nm laser excitation.
	\textbf{(e)} Optical images of a WS$_2$ homostructure achieved by the stacking of two thin-film crystals (thickness = 30 nm) with a 30$^\circ$ twist angle before (left panel) and after (right panel) fabrication of single nanopillars. The fabrication steps are the same as for \textbf{(c)}. Left inset: SEM image of a fabricated 30$^\circ$ twisted nanopillar. Right inset: AFM scan of a nanopillar in the 30$^\circ$ homostructure with hexagonal outlines to show the orientation of the top (black) and bottom (red) crystal. 
	\textbf{(f)} SHG spectrum collected from a 30$^\circ$ twisted nanopillar (r = 290 nm, h = 60 nm) under identical excitation as \textbf{(d)}. All AFM and SEM scale bars = 200 nm.}
  \label{F2}
 \end{figure}

\textbf{Nanoresonator fabrication} In order to demonstrate the ease of fabrication of layered materials in the fabrication of nanophotonic structures, we pattern nanopillars into many of the optically studied materials. After mechanical exfoliation of multilayered crystals of each material onto a nominally 290 nm SiO$_2$ on silicon substrate, we spun a positive resist onto the sample and patterned it into arrays of disks with varying radii using electron beam lithography (EBL). After development, the resist pattern was transferred into the layered crystals via reactive ion etching (RIE, see Methods) yielding monomer and dimer geometries. Figure \ref{F2}(a) displays a schematic representation of the fabrication process which results in differing geometries in some TMDs, namely circular and hexagonal as a result of an anisotropic and isotropic etch respectively. This is due to etch selectivity in the armchair as opposed to the zigzag crystal axis \cite{Zotev2022,Munkhbat2020,Munkhbat2022,Danielsen2021}.

It may be expected that all TMDs will result in similar geometries after an isotropic fabrication procedure due to their identical crystal symmetry, however, this is not observed as shown in Figure \ref{F2}(b). For a small nanopillar radius of r = 50 nm, WS$_2$, WSe$_2$ and MoS$_2$ result in a hexagonal geometry, however, this is not true for MoSe$_2$ which yields a circular geometry. For larger nanopillar radii, most of the materials also result in a circular geometry with the exception of WS$_2$ which maintains a hexagonal shape up to a radius of 300 nm. The transition from hexagonal to circular geometry can be seen at a radius of 150 nm for MoS$_2$ and 250 nm for WSe$_2$. As these crystals have undergone the same etching conditions for the same amount of time, this suggests that ionized fluorine radicals react with and remove some materials faster than others, leading to a hexagonal geometry in larger structures of certain TMDs \cite{Danielsen2021}. This is supported by theoretical calculations and experimental results for the enthalpy of formation of each material: -120.2 kJ/mol(WS$_2$,\cite{Chen2020a}), -185.3 kJ/mol(WSe$_2$,\cite{OHare1988}), -234.2 kJ/mol(MoSe$_2$,\cite{OHare1987}), -271.8 kJ/mol(MoS$_2$,\cite{OHare1988}), where the increasingly negative values indicate a more stable material. Thus, the less negative values suggest that dissociation of materials such as WS$_2$ and WSe$_2$ requires less energy than MoS$_2$ and MoSe$_2$. An additional contribution to the formation of a hexagonal geometry is the crystal thickness of each of the etched materials, which are as follows: 43 nm(WS$_2$), 85 nm(WSe$_2$), 78 nm(MoS$_2$), 29 nm(MoSe$_2$). A smaller thickness leads to less surface area for etching in plane and thus results in less hexagonal nanopillars. These two contributions of the crystal thickness and the enthalpy of formation suggest relative etching speeds of TMD materials which can be ranked as: WS$_2$ > WSe$_2$ > MoS$_2$ $\approx$ MoSe$_2$.

Fabrication of hBN and HfSe$_2$ nanopillars yields a circular geometry regardless of which previously employed etching recipes was used (see Supplementary Note 3) suggesting low or no crystal axis selectivity. Nanopillars in GaS also yield a circular geometry (Supplementary Note 3), although, the etching step was changed to employ a chlorine gas due to the poor reactivity of fluorine radicals with this material (see Methods). Similar to WS$_2$, fabrication of ZrSe$_2$ can yield both circular and hexagonal geometries up to a radius of 400 nm depending on the etching recipe used (Supplementary Note 3) suggesting a similarly fast etching speed and low enthalpy of formation. Subsequent attempts to etch In$_2$Se$_3$, NiPS$_3$ and MnPSe$_3$ with all etching recipes involving fluorine and chlorine gases yielded a very poor etch rate which did not form any appreciable nanostructures in the thin-film crystals (see Supplementary Note 3). Different methods of etching involving other gasses or wet etching may yield more satisfactory results, however, this is beyond the scope of this work.

In order to demonstrate the fabrication possibilities enabled by using vdW materials for nanophotonics, we also etched single nanopillars into twisted homostructures achieved by the stacking of two WS$_2$ thin-film crystals via an all-dry transfer technique (see Methods) available only to layered dielectrics. We fabricated nanopillars into two homostructures each of which consisted of two thin-film crystals stacked at 60$^\circ$ and 30$^\circ$ twist angles on SiO$_2$ substrates as shown in the left panels of Figure \ref{F2}(c) and (e) respectively. The thickness of the constituent crystals used in the fabrication of the homostructures was 50 nm and 30 nm for the 60$^\circ$ and 30$^\circ$ twist angle samples respectively. Utilizing EBL pattering and isotropic etching, we obtained single nanopillars with heights of 100 nm and 60 nm for the 60$^\circ$ and 30$^\circ$ twisted homostructures as shown in the right panels of Figure \ref{F2}(c) and (e) respectively. As the crystal axes of each constituent thin-film crystal in the homostructure are rotated with respect to each other due to the twist angle, the isotropic etching step will result in a rotated hexagonal structure for each. This rotation is not visible in the nanopillars of the 60$^\circ$ sample, shown in the insets of the right panel of Figure \ref{F2}(c), as the twist angle will yield hexagonal structures which form directly above each other. However, the 30$^\circ$ twist angle is visible in the nanopillars fabricated from the other homostructure shown in the insets of the right panel of Figure \ref{F2}(e) as this yields hexagonal structures with a 30$^\circ$ rotation. The twist angle in an unetched portion of the crystal was confirmed with SHG experiments, which yielded a brighter signal from the 60$^\circ$ homostructure when compared to the 30$^\circ$ sample (see Supplementary Note 4). This result is expected as the thickness of the WS2$_2$ homostructures was far below the coherence length of the SHG signal in this material and the centrosymmetry at the interface between the two constituent crystals was increasingly broken with twist angle \cite{Yao2021}. This same bright SHG signal was also observed for 60$^\circ$ nanopillar structures as opposed to those fabricated in the 30$^\circ$ homostructures (r = 290 nm for both homostructures) as shown by the spectra plotted in Figures \ref{F2}(d) and (f). This confirms that the broken centrosymmetry of the interface is consistent for both the unetched crystal and nanopillar structures as no additional enhancement observed from photonic resonances.\\

\begin{figure}[ht!]
	\centering
  \includegraphics[width=\linewidth]{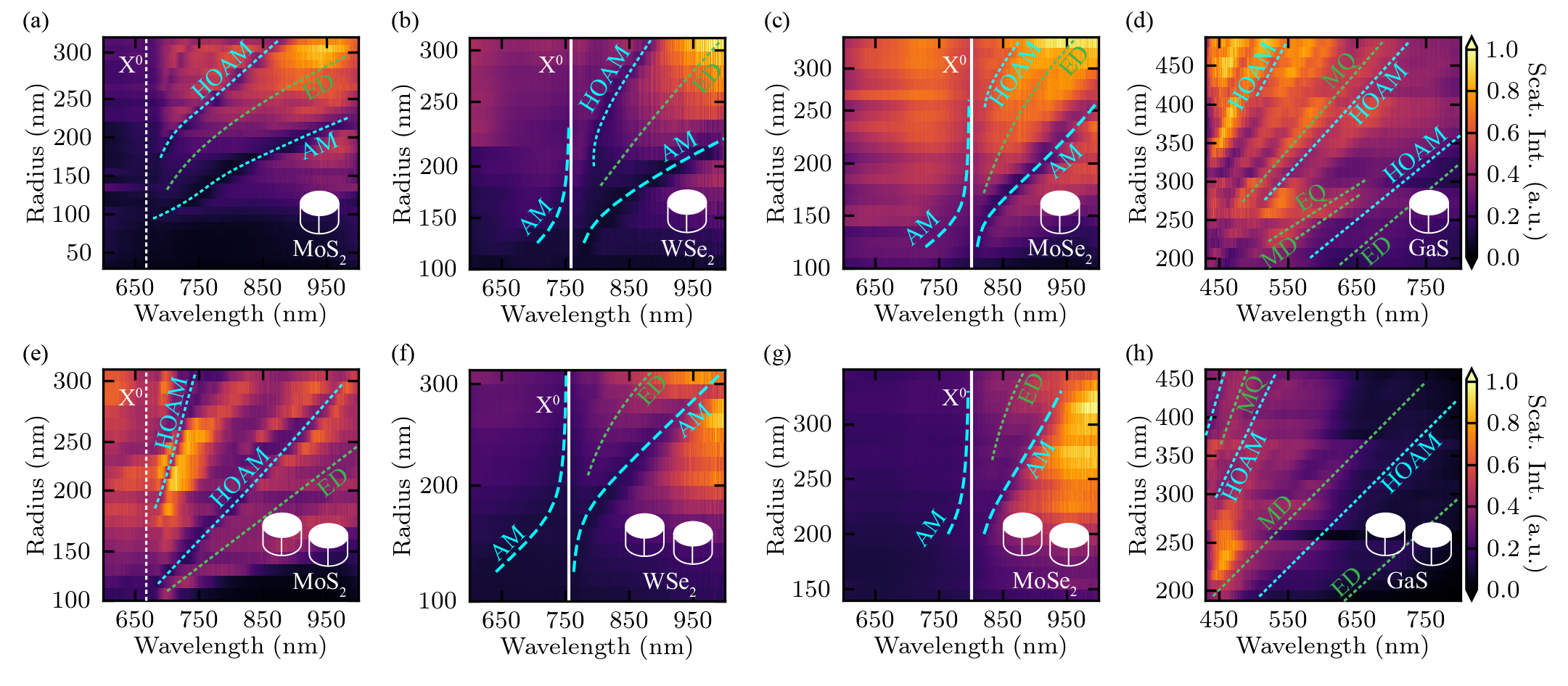}
  \caption{\textbf{Dark field spectra of monomer and dimer nanoantenna resonances in a variety of vdW materials.}
	\textbf{(a)}-\textbf{(d)} Experimental dark field spectra for monomer nanoantennas with a range of radii in MoS$_2$ (h = 70 nm), WSe$_2$ (h = 85 nm), MoSe$_2$ (h = 35 nm) and GaS (h = 200 nm). Identified resonances include the neutral exciton resonance of the material (X$^0$, white), the magnetic dipole (MD, green), electric dipole (ED), magnetic quadrupole (MQ) and electric quadrupole (EQ) resonances as well as the anapole mode (AM, cyan) and higher order anapole mode (HOAM). 
	\textbf{(e)}-\textbf{(h)} Experimental dark field spectra for dimer nanoantennas with a range of radii in MoS$_2$ (h = 190 nm), WSe$_2$ (h = 45 nm), MoSe$_2$ (h = 30 nm) and GaS (h = 145 nm). Dimer gaps range from 50 to 200 nm. Identified resonances are similar to those in monomer nanoantennas.}
  \label{F3}
\end{figure}

\textbf{Photonic resonances of fabricated nanoantennas} We subsequently studied the fabricated single and double nanopillar structures from a range of these materials using dark field spectroscopy. The height, radius and dimer separation gap were measured using AFM and SEM techniques. Figures \ref{F3}(a)-(d) show the dark field spectra of monomer nanopillars over a range of radii for different layered materials including MoS$_2$ (h = 70 nm), MoSe$_2$ (h = 35 nm), WSe$_2$ (h = 85 nm) and GaS (h = 200 nm). When compared with simulations of the scattering cross section expected for these nanostructures, we observed close agreement enabling us to identify neutral excitonic absorption in the material (X$^0$), geometric Mie resonances such as magnetic dipole (MD), electric dipole (ED), magnetic quadrupole (MQ) and electric quadrupole (EQ) resonances as well as light-confining anapole (AM) and higher order anapole (HOAM) modes (see Supplementary Note 5). Similarly, we also measured the dark field spectra of dimers in MoS$_2$ (h = 190 nm), MoSe$_2$ (h = 30 nm), WSe$_2$ (h = 45 nm) and GaS (h = 145 nm) with gaps ranging from 50 to 200 nm, shown in Figure \ref{F3}(e)-(h) and compared these to simulations (see Supplementary Note 5) which also agreed well, demonstrating the possibility of realizing more complex nanophotonic architectures. Additionally we have measured the dark field spectra of monomers and dimers fabricated from WS$_2$ and hBN which also yield similar resonances (see Supplementary Note 6). For GaS and hBN, we observe the formation of resonances at near-ultraviolet wavelengths due to the lack of absorption in these materials. We also observe a multitude of higher-order peaks in the dark field spectra of GaS monomer nanoantennas which are a result of a superposition of multiple Mie resonances, such as a magnetic and electric quadrupole as well as an electric dipole mode. Due to fabrication imperfections, some of the peaks are more or less prominent in the experimental than the simulated spectra. The WS$_2$ twisted homostructure monomers shown in Figure \ref{F2}(c) and (e) also yielded Mie and anapole resonances (see Supplementary Note 4) confirming them to be nanoantennas.

Focusing solely on the TMDs we observe an anticrossing of the anapole modes with the neutral exciton for TMD nanoantenna radii at which the energies of these two are expected to be degenerate providing evidence of strong light-matter coupling. As absorption is high at wavelengths below that of the X$^0$, the higher order anapole resonance is not as clearly defined in this range and its signature minimum is not clearly visible in the data limiting our ability to fit this. We have, however, fitted the anticrossing of the anapole mode with the neutral exciton for WSe$_2$ and MoSe$_2$ monomers using a coupled oscillator model. We extracted an energy splitting of 141 meV for WSe$_2$ monomers (r = 120 nm, h = 85 nm) as well as 100 meV for MoSe$_2$ monomers (r = 135 nm, h = 35 nm). Using a previously reported condition which indicates that the coupling strength of the exciton and photon resonances must be larger than the average of the individual linewidths \cite{Dovzhenko2018,Shen2022} (see Supplementary Note 5), the measured energy splittings in WSe$_2$ and MoSe$_2$ monomers were confirmed to be evidence of strong light-matter coupling. A similar anticrossing was observed for dimer structures in WSe$_2$ (132 meV for r = 155 nm, h = 45 nm), MoSe$_2$ (104 meV for r = 215 nm, h = 30 nm) and WS$_2$ (153 meV for r = 130 nm, h = 40 nm, see Supplementary Note 6). However, due to the larger linewidths of the neutral exciton and anapole resonances in these structures, the anticrossings do not satisfy the strong coupling condition instead indicating intermediate coupling by satisfying a complementary condition ($\Omega_{R}>(\frac{1}{\gamma_{X^0}} - \frac{1}{\gamma_{AM}})/2$) \cite{Shen2022}. FDTD simulations also confirm the strong and intermediate coupling between excitons and anapole modes yielding expected Rabi splitting as high as 200 meV for WSe$_2$ monomers and 179 meV for WS$_2$ dimers. Additionally, a higher order anapole mode is also observed to strongly couple to the exciton in simulations yielding Rabi splittings as high as 210 meV for WS$_2$ monomers. Smaller experimental anticrossings were also observed for monomer nanoantennas fabricated from WS$_2$ (see Supplementary Note 6) and MoS$_2$, which was limited due fabrication imperfections in these nanostructures as well as large absorption at wavelengths below the neutral exciton resonance limiting our ability to accurately fit these features. \\

\textbf{Dielectric nanoantennas on a metallic substrate} We have demonstrated that the fabrication of vdW nanoresonators on a low refractive index substrate such as SiO$_2$ is possible and results in well formed resonances in structures with heights as low as 30 nm. While this large index mismatch between substrate and nanoantenna may lead to tightly confined resonances, it can be advantageous to fabricate dielectric nanostructures onto a reflective substrate such as a gold mirror. This is expected to enable very high Q factor (10$^3$) dielectric-plasmonic modes which can provide very large Purcell factors (> 5000) \cite{Yang2017}. We thus proceed to fabricate an array of WS$_2$ monomer nanoantennas directly on a substrate with a 130 nm gold film. After exfoliation directly on the gold, similar to the technique used for a SiO$_2$/Si substrate, EBL and RIE is used to define WS$_2$ monomer nanoantennas with a circular and hexagonal geometry using the previously described recipes. An additional benefit to fabricating nanostructures in vdW materials onto a gold substrate arises due to the low etch rate of the gold compared to the previously used SiO$_2$. This forms a natural etch stop which allows for a higher tolerance in fabrication errors concerning the etch rate and time. 

\begin{figure}[ht!]
	\centering
  \includegraphics[width=\linewidth]{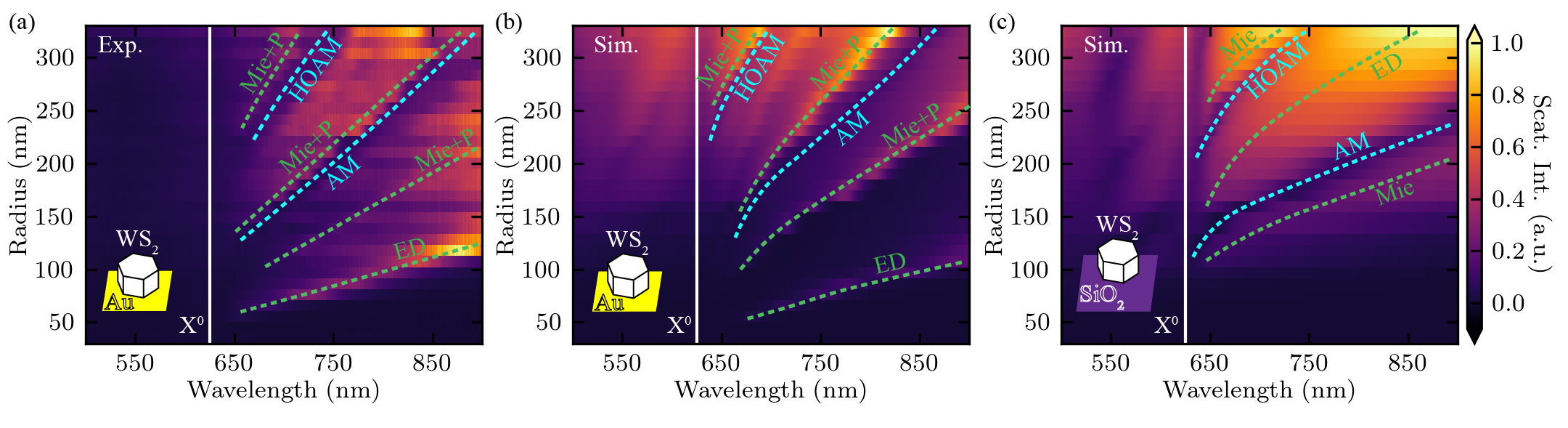}
  \caption{\textbf{Experimental and simulated dark field spectra of WS$_2$ hexagonal monomer nanoantennas on different substrates.}
	\textbf{(a)} Experimental dark field spectra of WS$_2$ hexagonal monomer nanoantennas (h = 62 nm) with a range of radii fabricated onto a susbrate with a 130 nm thick gold film. Identified resonances include the neutral exciton resonance of WS$_2$ (X$^0$, white solid line), Mie resonances such as the electric dipole, magnetic quadrupole and electric quadrupole resonances hybridized with a plasmonic component (Mie+P, green dashed curves) as well as the anapole mode (AM, cyan dashed curves) and higher order anapole mode (HOAM). 
	\textbf{(b)} Simulated scattering intensity of WS$_2$ hexagonal monomer nanoantennas with the same height and range of radii as in \textbf{(a)} on a semi-infinite gold substrate showing close agreement to the experimental results. Identified resonances are the same as those shown in \textbf{(a)}.
	\textbf{(c)} Simulated scattering intensity of WS$_2$ hexagonal monomer nanoantennas with the same height and range of radii as in \textbf{(a)} on a semi-infinite SiO$_2$ substrate displaying broadened resonances. Identified resonances include similar yet broadened Mie as well as anapole and higher order anapole modes.}
  \label{F4}
\end{figure}

We subsequently record the dark field spectra of an array of the fabricated hexagonal WS$_2$ monomer nanoantennas on gold, plotted in Figure \ref{F4}(a). After characterization of the height (h = 62 nm) and radii of the nanoantennas using AFM and SEM respectively, we simulate the expected scattering intensity, shown in Figure \ref{F4}(b) and observe close agreement with experiment. This allows us to identify Mie and anapole modes similar to those found in the nanostructures fabricated on SiO$_2$. 

In order to compare these structures to those previously fabricated on a low refractive index substrate, we also simulate the scattering intensities of identical WS$_2$ hexagonal monomer nanoantennas onto a SiO$_2$ substrate, displayed in Figure \ref{F4}(c). Similar resonances are identified in these nanostructures, however, we observe a blueshifted ED mode and redshifted anapole modes which appear broader than those recorded for a gold substrate. For larger radii, we observe the appearance of additional modes in the nanoantennas fabricated on a gold substrate which we attribute to the formation of hybrid Mie-plasmonic (Mie+P) resonances. A plasmonic contribution may also be present for smaller nanoantenna radii, hybridizing with and enhancing spectral features which we identified as dielectric Mie (ED) and anapole resonances \cite{Yang2017} due to mirror charge currents in the gold which respond to the induced bound charge currents in the WS$_2$ nanoantennas. The reduced broadening of resonances in the nanostructures fabricated on gold leads us to consider the confinement of the electric field inside and in close proximity to the nanoantenna structure. Electric field intensity profiles recorded in and surrounding a WS$_2$ monomer nanoantenna at the anapole and electric dipole modes for a gold and SiO$_2$ substrate reveal differing confinements, shedding light on the origin of the mode shift, as well as increased intensities by at least an order of magnitude (see Supplementary Note 7). These suggest that the contribution of plasmonic resonances to the observed features in scattering may not be negligible. 

Additionally, we compare the quality factors of the identified ED resonance in monomer nanoantennas fabricated from different vdW materials on a gold and SiO$_2$ substrate both in simulation and experiment as shown by Table \ref{tab1}. The ED resonance was chosen as it was observed in all of the experimentally studied nanoantennas and leads to a high outcoupling of emitted light, which is important for most nanophotonic applications. The black and orange numbers correspond to a SiO$_2$ and gold substrate respectively. Each value was extracted from a lorentzian fit of a resonance peak in scattering for a monomer nanoantenna geometry yielding a mode far from any anticrossings. For all materials, the extracted quality factors of resonances in nanostructures on a gold substrate yield higher values than for a SiO$_2$ substrate in both experiment and simulation as suggested by the large reduction in broadening observed in Figure \ref{F4}(a) and (b) when compared to Figure \ref{F4}(c). Previously reported Mie resonances in dielectric monomer nanoantennas \cite{Miroshnichenko2015,Timofeeva2018} also yield lower quality factors than observed in our vdW nanoantennas on gold. These results suggest that the hybridization of Mie and plasmonic resonances may provide a large contribution to the quality factor improvement. 

Some applications may require the placement of nanophotonic structures onto substrates which can be damaged by standard nanofabrication techniques such as biological samples \cite{Xu2018} or substrates with deposited monolayer TMDs. We show that the nanoantennas we fabricate can be transferred onto such sensitive surfaces by taking advantage of their weak van der Waals adhesion to the substrate. Supplementary Note 8 shows the transfer of WS$_2$ hexagonal nanoantennas onto a monolayer and bilayer WSe$_2$ crystal which has been previously deposited onto a gold substrate. This regime of nanostructure fabrication, which we name "transferable photonics", enables the formation of tightly confined hotspots which can provide many orders of magnitude enhancement to emission sources, previously demonstrated for plasmonic structures \cite{Chikkaraddy2016}, as well as applications which contain sensitive samples that may be damaged by standard nanofabrication techniques. Further development of this method is required in order to achieve a more controlled pick up and transfer of the nanostructures, which may prove easier for other nanoresonators such as photonic crystal cavities or metasurfaces \cite{Liu2018c}, however, this is beyond the scope of this work.\\

\begin{figure}[ht!]
	\centering
  \includegraphics[width=\linewidth]{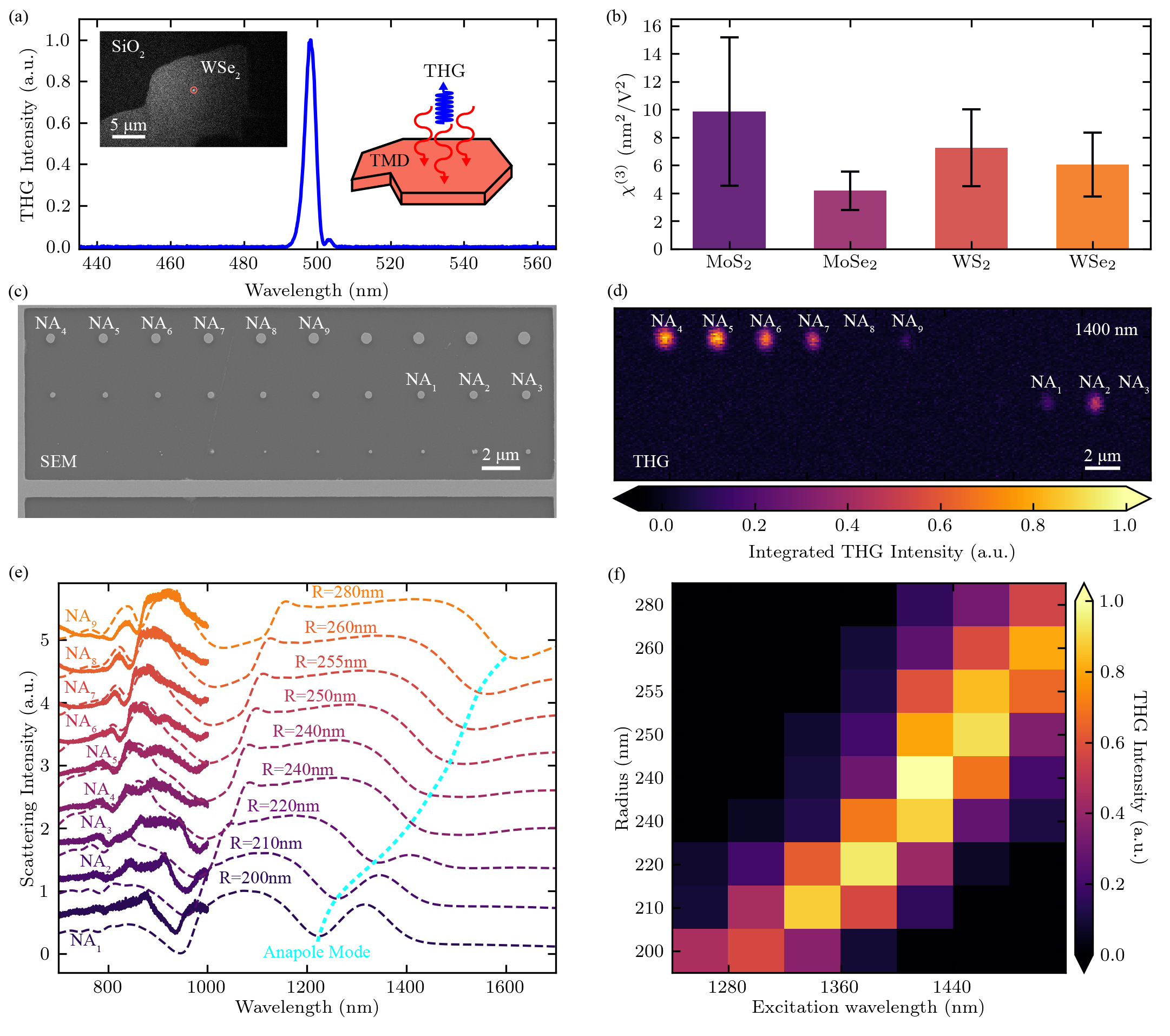}
  \caption{\textbf{Third harmonic generation experiments with TMDs.}
	\textbf{(a)} Third harmonic generation spectrum from WSe$_2$ thin-film crystal under 1494 nm excitation. Left inset: Optical image of a WSe$_2$ crystal under study. Right inset: schematic of third harmonic generation from TMD crystals.
	\textbf{(b)} Third harmonic susceptibilities ($\chi^{(3)}$) extracted for different TMD crystals for 1500 nm excitation. The error bars show the error in measurements of the thickness via AFM.
	\textbf{(c)} SEM image of array of WSe$_2$ monomers under study with labels for the nanoantennas of interest. 
	\textbf{(d)} Map of the THG signal across the array shown in \textbf{(c)} under 1400 nm excitation. Nanoantennas NA$_4$ and NA$_5$ exhibit the brightest THG signal for this wavelength of illumination. 
	\textbf{(e)} Dark field spectrum of each monomer nanoantenna compared with simulations of the scattering cross section of each geometry. The formation of an anapole, which can be used for third harmonic generation enhancement, is highlighted.
	\textbf{(f)} THG signal intensity from each nanoantenna under a range of excitation wavelengths. A resonant enhancement is observed at larger nanoantennas for increasing excitation wavelengths as expected from the redshifting anapole mode to which the THG signal is coupled.}
  \label{F5}
\end{figure}

\textbf{Nonlinear optical properties and applications} As there is a large interest in employing vdW materials for enhancing nonlinear light applications \cite{Busschaert2020,Popkova2022,Zotev2022,Xu2022}, we asses the viability of the most widely used TMDs for higher harmonic generation by exciting several thin-film crystals from each material with 220 fs laser pulses, nominally centered at 1500 nm. Figure \ref{F5}(a) displays the third harmonic generation spectrum from a WSe$_2$ thin-film crystal. The THG signal is observed at 498 nm (as the excitation has drifted to 1494 nm), with a small shoulder that we attribute to the laser spectrum, which also exhibits this feature. We observe a strong THG signal from this and other TMD materials and thus we extract the third harmonic susceptibility ($\chi^{(3)}$) for each by comparing our recorded THG intensity with nonlinear scattering theory \cite{OBrien2015} coupled with transfer-matrix method simulations (see Supplementary Note 9). The extracted $\chi^{(3)}$ values are plotted in Figure \ref{F5}(b) where the error bars indicate the uncertainty in the measurement of the thickness of each thin-film crystal via AFM. The third harmonic susceptibilities we extract are up to an order of magnitude higher than for monolayers of the same material \cite{Wang2022a,Rosa2018}, which is expected due to the difference in nonlinear properties between monolayer and multilayer crystals as well as the strong thickness dependence of the model used to to extract the $\chi^{(3)}$ values. These values are also up to 4 orders of magnitude larger than in BBO crystals regularly used for laser frequency tripling \cite{Penzkofer1989}. We observe the highest third harmonic susceptibility in MoS$_2$ and the lowest in MoSe$_2$.

Next, we pattern monomer nanoantennas with varying radii (r = 200 - 280 nm) into a WSe$_2$ crystal with 250 nm thickness. An SEM image of the fabricated array is displayed in Figure \ref{F5}(c). We record the dark field spectra of the nanoantennas in the visible range and compare these to simulations including both visible and near-infrared wavelengths which are in good agreement as shown in Figure \ref{F5}(e). From the simulated infrared scattering intensity we observed an anapole mode scattering minimum redshifting from 1200 to 1600 nm with increasing nanoantenna radius, which can be used for enhancing higher harmonic generation. Thus, we illuminate the nanoantennas with a laser at a wavelength of 1400 nm and reposition our excitation and collection spot across the array of nanoantennas in order to record a map of the THG signal (see Methods). As shown in Figure \ref{F5}(d), the THG signal is maximized at the position of certain nanoantennas (r = 240 nm) as opposed to others indicating a resonant coupling of the excitation laser leading to higher third harmonic signal. The nanoantennas which exhibit the brightest THG signal at 1400 nm illumination also exhibit an anapole mode minimum near this wavelength, as shown in Figure \ref{F5}(e), providing further evidence for coupling of the excitation to the resonance of the nanoantenna. We subsequently vary the wavelength of the illumination source and observe the maximum THG signal shift from lower radius nanoantennas to those with a higher radius for an increasing excitation wavelength as shown in Figure \ref{F5}(f) thereby confirming that the excitation is coupling to the anapole mode in the WSe$_2$ monomer nanoantennas.\\

\large
\textbf{Discussion}\\
\normalsize

We have studied a range of different vdW materials for their potential use as nanophotonic structures and demonstrated their advantages compared to traditional dielectric or plasmonic materials. We extract the dielectric response of 11 different layered materials via micro-ellipsometry. We observe very high refractive indices ($n > 5$ for some materials) when compared to Si or GaAs ($n \approx 4$ \cite{Aspnes1983}) as well as a range of transparency windows from ultra-violet to near-infrared wavelengths. Utilizing material specific fitting models to extract the linear optical parameters leads to very high values of birefringence ($\Delta n \approx$ 3.0) in MoS$_2$ and ZrSe$_2$ as well as transparency for light propagating in-plane with an out-of-plane electric field in WS$_2$, WSe$_2$, MoS$_2$, MoSe$_2$, ZrSe$_2$, HfSe$_2$, GaS and hBN. 

We fabricate nanoantenna structures in widely used vdW materials (MoS$_2$, MoSe$_2$, WS$_2$, WSe$_2$, hBN, HfSe$_2$, ZrSe$_2$, GaS). The geometries of nanostructures undergoing the same isotropic fabrication conditions provide insight into the etching speed of some TMDs which can be ranked as: ZrSe$_2$ $\approx$ WS$_2$ > WSe$_2$ > MoSe$_2$ $\approx$ MoS$_2$. Etching of In$_2$Se$_3$, MnPSe$_3$ and NiPS$_3$ nanoantennas is beyond the scope of this work, however, our attempts with fluorine and chlorine gasses using both isotropic and anisotropic conditions indicate that other RIE or wet etching approaches will be necessary. Nanoantenna fabrication in MnPSe$_3$ and NiPS$_3$ may lead to optical control of the magnetic properties of these materials via coupling to a magnetic dipole mode. Additionally, our demonstration of monomer nanoantennas in twisted WS$_2$ homostructures provides a straightforward route for future fabrication of heterostructure nanophotonic resonators such as 3D photonic crystal cavities or moire architectures \cite{Nguyen2022} with integrated emissive materials, such as monolayer TMDs, enabled by the the weak van der Waals adhesion of layered dielectrics. The SHG signal observed from the interface of the twisted WS$_2$ homostructure, with a thickness far below the coherence length in this material, can be enhanced due to coupling with the anapole mode in nanoantennas and may provide insights into interlayer excitons in multilayer TMD crystals.

Dark field spectroscopy of the fabricated structures in MoS$_2$, MoSe$_2$, WS$_2$, WS$_2$ twisted homostructures, WSe$_2$, GaS and hBN on SiO$_2$ yields well defined Mie and anapole resonances from ultra-violet to near-infrared wavelengths. Strong coupling within single nanostructures was observed at room temperature in several TMD crystals with extracted Rabi splittings as high as 153 meV for WS$_2$, 141 meV for WSe$_2$ and 104 meV for MoSe$_2$ nanoantennas. This is a factor of 5 larger than reported for monolayer TMDs in high Q microcavities \cite{Dufferwiel2018,Gillard2021} and 1 to 2 orders of magnitude higher than in InGaAs \cite{Skolnick1998}, AlGaAs \cite{Savona1995,Suchomel2017} and AlGaN \cite{Christmann2008} multiple quantum wells in similar microcavities. 

Evidence of the possibility of fabricating vdW nanostructures on virtually any substrate without the need for lattice matching is provided by the patterning of nanoantennas onto a SiO$_2$ as well as gold substrate. This provides the possibility for straightforward realization of large refractive index contrast interfaces as well as the integration of dielectric and plasmonic nanophotonic devices \cite{Yang2017}. The fabricated WS$_2$ nanoantennas on gold yield highly confined resonances with improved Q factors which may be beneficial for a number of applications including Purcell enhancement of emission, quantum efficiency enhancement \cite{Yang2017}, collection efficiency enhancement, strong light matter coupling, optical trapping, and surface enhanced Raman spectroscopy among others. Additional possibilities, demonstrated by our deposition of pre-fabricated WS$_2$ nanoantennas onto a WSe$_2$ monolayer on a gold substrate, include the transfer of etched nanostructures from one substrate onto another which contains sensitive samples such as monolayer TMDs or biological samples.

We also characterize the nonlinear optical properties of some of the most widely studied vdW crystals by extracting the THG susceptibility of various thin-film TMDs near the telecom C band, advantageous for quantum applications \cite{Bovino2009}, yielding values up to an order of magnitude higher than in monolayers \cite{Wang2022a,Rosa2018} and up to 4 orders of magnitude larger than in BBO crystals regularly used for laser frequency tripling applications \cite{Penzkofer1989}. We also explore a method of enhancing such nonlinear signals via coupling to resonances in monomer nanoantennas of WSe$_2$.

Due to their high refractive indices, wide range of bandgaps and adhesive properties, vdW material nanostructures enable a wide variety of applications as shown by our demonstrations of twisted homostructure nanostructures, single nanoantenna strong coupling, hybrid high-Q Mie-Plasmonic modes, post-fabrication nanostructure deposition and THG enhancement. The advantages of using van der Waals materials will enable many exciting opportunities in nanophotonics.\\

\large
\textbf{Methods}\\ 
\normalsize

\textbf{Ellipsometry}
Spectroscopic ellipsometry measurements were carried out in the wavelength range 360 nm to 1000 nm with a spatial resolution of $\approx$ 1 $\mu$m$^2$ using a nulling imaging ellipsometer (EP4, Accurion Gmbh) in an inert Ar atmosphere at room temperature. Ellipsometric data from the samples were acquired at three different angles of incidence (AOI, defining the vertical as AOI = 0$^\circ$) at 45$^\circ$, 50$^\circ$ and 55$^\circ$.

\textbf{Sample fabrication}
\textit{Van der Waals materials exfoliation}: Layered material crystals were mechanically exfoliated from bulk (HQ-graphene) onto a nominally 290 nm SiO$_{2}$ on silicon or gold substrate. Large crystals with recognizable axes via straight edged sides at 120$^{\circ}$ to each other were identified and their positions within the sample were recorded for further patterning.

\textit{Homostructure fabrication}: The 60$^\circ$ and 30$^\circ$ stacked homostructures were fabricated with an all dry transfer technique. For the 60$^\circ$ homostructure, two separate 50 nm WS$_2$ crystals were exfoliated onto a PPC/SiO$_2$ substrate. These were then picked up consecutively with a 60$^\circ$ twist angle using a PMMA/PDMS membrane and deposited onto a SiO$_2$ substrate. For the 30$^\circ$ homostructure, a single 30 nm WS$_2$ crystal was exfoliated onto a PPC/SiO$_2$ substrate and subsequently broken with the use of an AFM cantilever tip. Consecutive pick up of the two crystals with a 30$^\circ$ twist angle using a PMMA/PDMS membrane was followed by deposition onto another SiO$_2$ substrate. Thicknesses were measured with AFM. 

\textit{Electron beam lithography}: Samples were spin coated with ARP-9 resist (AllResist GmbH) at 3500 rpm for 60 s and baked at 180$^\circ$ for 5 min yielding a film of 200 nm thickness. Electron beam lithography was performed in a Raith GmbH Voyager system operating at 50 kV using a beam current of 560 pA.

\textit{Reactive ion etching of TMDs and hBN}: Anisotropic etching to imprint the resist pattern into the WS$_{2}$ crystals physically was carried out using a mixture of CHF$_{3}$ (14.5 sccm) and SF$_{6}$ (12.5 sccm) at a DC bias of 180 V and a pressure of 0.039 mbar for 40 seconds. Isotropic etching was achieved by using a more chemical recipe with solely SF$_{6}$ (30 sccm) at a DC bias of 40 V and a pressure of 0.13 mbar for 40 seconds. Removal of the remaining resist after etching was accomplished by a bath in warm 1165 resist remover (1 hour) followed by Acetone (5 min) and IPA (5 min). If resist is still found on the sample, final cleaning is done in a bath of Acetone (1 hour) and IPA (5 min) followed by 1 hour in a UV ozone treatment. In some cases, the structures were slightly over-etched leading to nanoantennas with a small pedestal of SiO$_2$ (<20 nm) or gold (<5 nm). This, however, did not lead to any noticeable changes in the photonic resonances.

\textit{Reactive ion etching of GaS}: Isotropic etching of GaS was achieved with SiCl$_4$ gas (5 sccm) at a pressure of 50 mTorr and DC bias of 175 V for 7 minutes. The resist removal step was the same as for other materials, however, this did not achieve proper removal of all resist from the sample. The residual resist did not noticeably impact the photonic resonances measured in dark field spectroscopy.

\textit{Gold substrate preparation}: In order to prepare the gold substrate, we firstly deposit a 10 nm layer of Ti onto a 290nm SiO$_2$/Si substrate via e-beam evaporation in order to improve the adhesion between substrate and gold. We subsequently deposit 130 nm of gold via the same method.

\textbf{Dark field spectroscopy}
Optical spectroscopy in a dark field configuration was achieved using a Nikon LV150N microscope with a fiber-coupled output. Incident illumination from a tungsten halogen lamp in the microscope was guided to a circular beam block with a diameter smaller than the beam diameter. The light was then reflected by a 45$^\circ$ tilted annular mirror towards a 50x Nikon (0.8 NA) dark-field objective which only illuminates the sample at large angles to the normal. Reflected light from the sample is guided back through the same objective towards a fiber coupler. Due to the small diameter of the multimode fiber core used, only light reflected back at small angles to the normal is collected. The fiber from the microscope was subsequently coupled to a Princeton Instruments spectrometer and charge coupled device.

\textbf{FDTD scattering simulations}
Calculations of the scattering cross section shown in Figure \ref{F4}(c), \ref{F5}(e) and Supplementary Notes 5 and 6 were carried out by defining the geometry of the vdW material nanoantennas onto a SiO$_{2}$ or gold substrate utilizing the refractive indices extracted from the ellipsometry measurements. Illumination with a plane wave was sent normal to the surface using a TFSF source from the air side. The illumination was polarized parallel to the surface. The scattered intensity was subsequently collected from a monitor set above the illumination plane (in the far field) so that the dark field spectroscopy experiments could be closely emulated. The finite-difference time-domain simulations were carried out using Lumerical Inc. software.

\textbf{Second Harmonic Generation}
In order to probe the second harmonic generated signal from the twisted homostructures and their nanopillars, as shown in Figure \ref{F2}(d),(f) and Supplementary Note 4, we used a Mai-Tai Ti-sapphire mode-locked femtosecond laser as the excitation source set at 850 nm with an average power of 10 mW. The collimated laser light passed through a linear film polarizer, half wave plate, dichroic mirror and was incident on a 100x (0.7 NA) Mitutoyo objective which focused the excitation light onto the sample allowing us to probe single nanopillars and thin-film crystals. Second harmonic generated light was then collected using the same objective subsequently reflecting the light from the dichroic mirror and passing it through an analyzer. The collected light is then filtered by long-pass filters (650 nm cutoff) and fiber coupled to a multi-mode fiber and sent to a Princeton Instruments spectrometer and CCD to yield the data displayed in Figure \ref{F2}(d) and (f) as well as in Supplementary Note 4.

\textbf{Third Harmonic Generation}
THG measurements, shown in Figure \ref{F5}, were carried out by illuminating the sample with 220 fs laser pulses (Yb:KGW Pharos, Light conversion) at 100 kHz repetition rate and a wavelength of 1500 nm through a 0.85 NA objective (60X). The laser beam was directed towards the sample via a dichroic mirror which transmits the visible THG signal at 500 nm and reflects the illuminating wavelength. The illuminating beam is subsequently focused onto the back focal plane of the objective using a 1X telescope (100 mm) and the power is attenuated through a rotational neutral density wheel. The excitation power is measured with an infrared power meter (Thorlabs, S122C). The collected THG signal intensity from the sample is measured via a Thorlabs camera (CS165MUM) calibrated to convert counts/pixel to an intensity. The beam size of the THG signal and the excitation beam is measured by fitting the camera counts from a reference gallium phosphide sample emitting a strong second harmonic signal with a Gaussian function. The final peak intensity of the excitation and third harmonic signal is calculated by adjusting the power measurements with the respective beam sizes.

Nanoantenna THG resonances were recorded using a piezoelectric stage to scan the sample over the region of interest while the generated signal was recorded with a single-photon detector (Picoquant, Micro Photon Devices). The excitation wavelength ranged from 1240 nm to 1480 nm with increments of 40 nm using an excitation power of 6.5 $\mu$W. The reported THG intensity from each nanoantenna was determined using the counts of the single-photon detector and corrected with the beam size measurement so that excitation intensity fluctuations in this spectral range were taken into account. \\

\large
\textbf{Acknowledgments}\\
\normalsize

P. G. Z., T.S.M., S.R., X.H., and A. I. T. acknowledge support from the European Graphene Flagship Project under grant agreement number 881603 and EPSRC grants EP/S030751/1, EP/V006975/1, and EP/V026496/1. L.S. acknowledges funding support through a Humboldt Research Fellowship from the Alexander von Humboldt Foundation. Y.W. acknowledges a Research Fellowship (TOAST) awarded by the Royal Academy of Engineering. We would also like to thank Prof. Keith McKenna for his contribution concerning the enthalpy of formation factor in the etching speed of different vdW materials.\\

\large
\textbf{Author Contributions}\\
\normalsize

P.G.Z and T.S.M. exfoliated WS$_2$, WSe$_2$, MoSe$_2$, MoS$_2$, HfSe$_2$, ZrSe$_2$ hBN, GaS, In$_2$Se$_3$, MnPSe$_3$ and NiPS$_3$ crystals onto SiO$_2$ substrates for ellipsometry measurements, nanoresonator fabrication and THG susceptibility measurements. D.A.P. and M.B.G performed ellipsometry measurements and fit the complex reflectance ratio with appropriate analytical models in order to extract the real and imaginary parts of the complex refractive index of each material. Y.W., X.H. and P.G.Z. fabricated nanoantenna structures using EBL and RIE. P.G.Z. performed SHG experiments to confirm the twist angle of the homostructures and their nanoantennas. P.G.Z., T.S.M. and S.R. performed all AFM characterization of crystal thickness and nanoantenna heights. P.G.Z., T.S.M. and S.R. performed all SEM measurements to characterize nanoantenna radii. P.G.Z., T.S.M. and S.R. recorded all all dark field spectra and analyzed the results. C.L. and P.G.Z. fit the experimental and simulated scattering data to a coupled oscillator model to provide evidence of strong coupling and extract Rabi splittings. P.G.Z. and T.S.M. performed all FDTD simulations of the scattering cross sections. P.G.Z. and X.H. transferred WS$_2$ nanoantennas onto a monolayer WSe$_2$ crystal on a gold substrate. T.H. and S.V. performed THG experiments on WSe$_2$ nanoantennas as well as TMD thin-film crystals and subsequently analyzed the recorded spectra in order to extract $\chi^{(3)}$ values. T.F.K., B.D.G., R.S. and A.I.T. managed various aspects of the project. P.G.Z. and A.I.T. wrote the manuscript with contributions from all co-authors. P.G.Z., L.S., Y.W. and A.I.T. conceived the experiments and simulations. A.I.T. oversaw the entire project.\\

\bibliographystyle{unsrt}
\bibliography{./library}
\end{multicols}

\pagebreak

\begin{figure}[ht!]
	\centering
  \includegraphics[width=\linewidth]{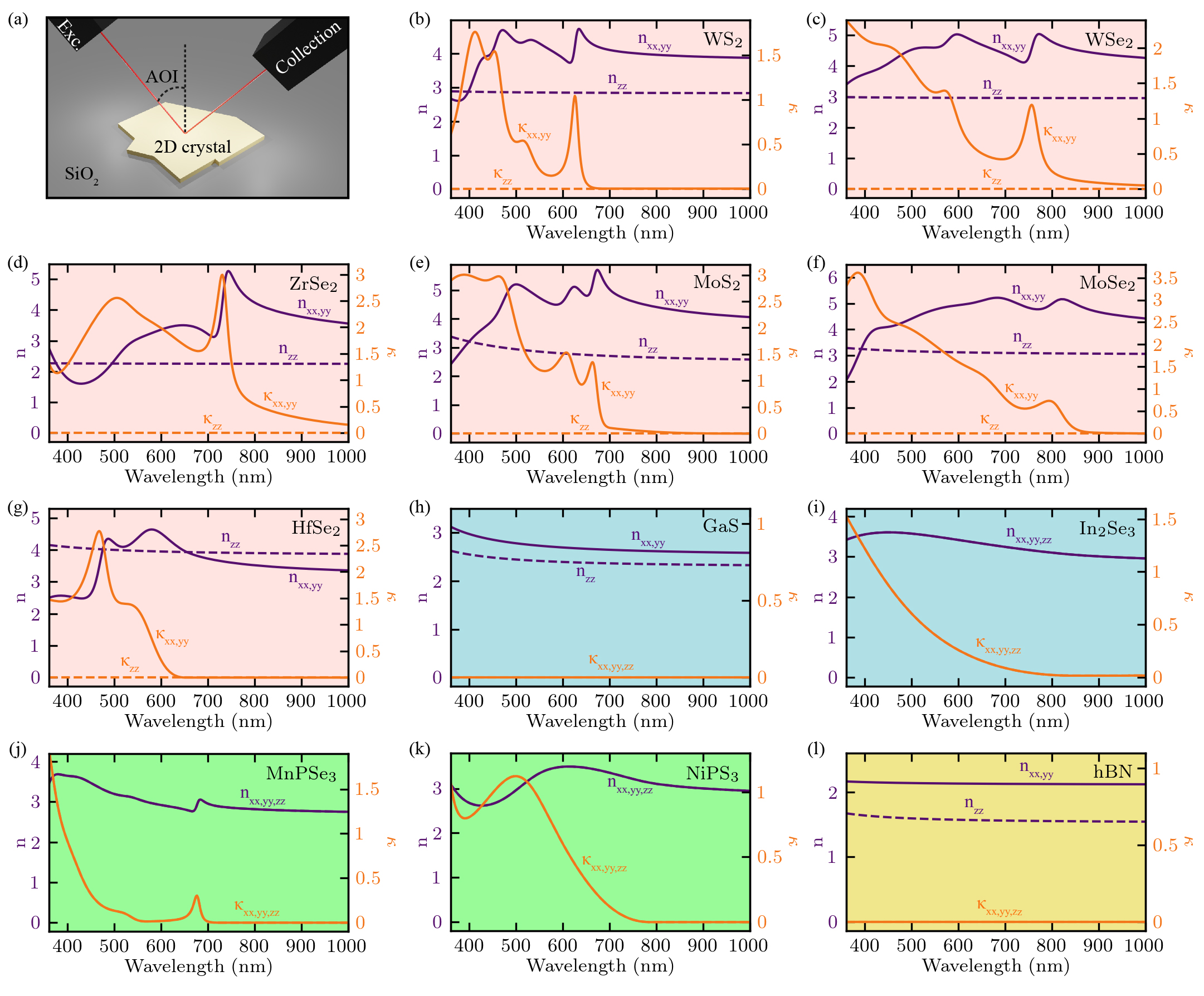}
  \caption{\textbf{Refractive indices and extinction coefficients of layered materials}
	\textbf{(a)} Graphical representation of our micro-ellipsometry setup used to record the complex reflectance ratio from which we extract the dielectric response of each material.
	Refractive indices ($n$) and extinction coefficients ($\kappa$) extracted from ellipsometry measurements of the layered materials. These include TMDs (faint red background): \textbf{(b)} WS$_2$, \textbf{(c)} WSe$_2$, \textbf{(d)} ZrSe$_2$, \textbf{(e)} MoS$_2$, \textbf{(f)} MoSe$_2$ and \textbf{(g)} HfSe$_2$; III-VI materials (blue background): \textbf{(h)} GaS and \textbf{(i)} In$_2$Se$_3$; magnetic materials (green background): \textbf{(j)} MnPSe$_3$ and \textbf{(k)} NiPS$_3$; insulating (yellow background) \textbf{(l)} hBN.}
  \label{F1}
 \end{figure}

\FloatBarrier

\FloatBarrier

\FloatBarrier

\FloatBarrier

\begin{table}[ht!]
\centering
\begin{tabular}{|c||c|c|}
\hline
 & \textit{\textbf{Experimental ED}} & \textit{\textbf{Simulated ED}} \\
\hline
\hline
WS$_2$ & 4.95$\pm$0.34 {\color{orange}(12.64$\pm$0.07)} & 3.99$\pm$0.05 {\color{orange}(14.27$\pm$0.06)} \\
\hline
WSe$_2$ & 4.66$\pm$0.26 & 6.58$\pm$0.10 {\color{orange}(11.72$\pm$0.05)} \\
\hline
MoS$_2$ & 4.04$\pm$0.03 & 4.63$\pm$0.05 {\color{orange}(14.94$\pm$0.05)} \\
\hline
MoSe$_2$ & 4.03$\pm$0.17 & 5.36$\pm$0.13 {\color{orange}(17.98$\pm$0.07)} \\
\hline
GaS & 5.21$\pm$0.11 & 5.26$\pm$0.49 {\color{orange}(6.11$\pm$0.06)} \\
\hline
Si \cite{Miroshnichenko2015} & 7.04$\pm$0.07 & \\
\hline
GaAs/AlGaAs \cite{Timofeeva2018} & & 4.60$\pm$0.04 \\
\hline
\end{tabular}
\caption{\textbf{Experimental and simulated quality factors for Mie resonators.} The quality factors were extracted from Lorentzian fits to experimental and simulated electric dipole resonances in different materials on different substrates. Values were extracted for resonators from this work as well as previous reports of monomer nanoantennas. Black and orange numbers represent quality factors extracted for monomer nanoantennas on a dielectric and gold substrate respectively.}
\label{tab1}
\end{table}

\FloatBarrier

\end{document}


\onehalfspacing
\pagenumbering{arabic}

\begin{center}
{\Large{\bf {\Large{\bf Supplementary Information for: Van der Waals Materials for Applications in Nanophotonics}}}}
\vskip1.0\baselineskip{Panaiot G. Zotev$^{a*}$, Yue Wang$^b$, Daniel Andres-Penares$^c$, Toby Severs Millard$^a$, Sam Randerson$^a$, Xuerong Hu$^a$, Luca Sortino$^d$, Charalambos Louca$^a$, Mauro Brotons-Gisbert$^c$, Tahiyat Huq$^e$, Riccardo Sapienza$^e$, Thomas F. Krauss$^b$, Brian Gerardot$^c$, Alexander I. Tartakovskii$^{a**}$}
\vskip0.5\baselineskip\footnotesize{\em$^a$Department of Physics and Astronomy, University of Sheffield, Sheffield, S3 7RH, UK\\$^b$School of Physics, Engineering and Technology, University of York, York, YO10 5DD, UK\\$^c$School of Engineering and Physical Sciences, Heriot-Watt University, Edinburgh, EH14 4AS, UK\\$^d$Chair in Hybrid Nanosystems, Nanoinstitute Munich, Faculty of Physics, Ludwig-Maximilians-Universit\"{a}t, 80539, Munich, Germany\\$^e$The Blackett Laboratory, Department of Physics, Imperial College London, London, SW7 2BW, UK\\}
$^{*}$p.zotev@sheffield.ac.uk \quad $^{**}$a.tartakovskii@sheffield.ac.uk

\end{center}
\vskip1.0\baselineskip


\setcounter{figure}{0}
\renewcommand{\thefigure}{S\arabic{figure}}

\begin{center}
\Large
\textbf{Supplementary Note 1: Determination of linear optical constants}
\end{center}

In order to determine the linear optical constants such as the refractive index and the extinction coefficient of each material, we employ spectroscopic ellipsometry on exfoliated thin-film crystals of known thickness. We record the amplitude ($\Psi$) and phase difference ($\Delta$) of the complex reflectance ratio at three different angles of incidence (45$^\circ$,50$^\circ$,55$^\circ$, AOI) for each material which are plotted below in Figures \ref{Ell1}-\ref{Ell3}. The experimental data points (open circles) are fitted with different analytical models (dashed curves) in order to extract the complex refractive index from each material. The results shown in Figures \ref{Ell1}-\ref{Ell3} are grouped according to the model used to fit the experimental data. In order to minimize substrate-induced uncertainties in the determination of the complex refractive indices, ellipsometric data was obtained from the bare SiO$_2$/Si substrate immediately adjacent to each studied layered material. A measure of the thickness of the SiO$_2$ layer was extracted by fitting this data with reported refractive indices of Si and SiO$_2$. 

\textbf{Large bandgap materials} The first group of materials includes GaS and hBN which both exhibit anisotropic crystal structures yet their bandgap is expected to be larger than the energies sampled by our experimental range (<360 nm). In this case, extinction coefficients can be neglected and we extract real-valued refractive indices described by a Sellmeier dispersion law of the form:

\begin{equation}
\begin{array}{lr}
n_{i}^{2}(\lambda) = 1 + \frac{B_{i}\lambda^{2}}{\lambda^{2}-C_{i}}, \\
k_{i}^{2}(\lambda) = 0,
\end{array}
\label{eq:1}
\end{equation}

\noindent
where $B_i$ and $C_i$ are the Sellmeier coefficients while $i=IP,OP$. In our analysis we extract a different index of refraction in-plane (IP) and out-of-plane (OP) as expected from the inherent anisotropy of the layered crystals. The ellipsometry data fitted with the Sellmeier dispersion law is plotted in Figure \ref{Ell1}.

\begin{figure}[ht!]
	\centering
  \includegraphics[width=\linewidth]{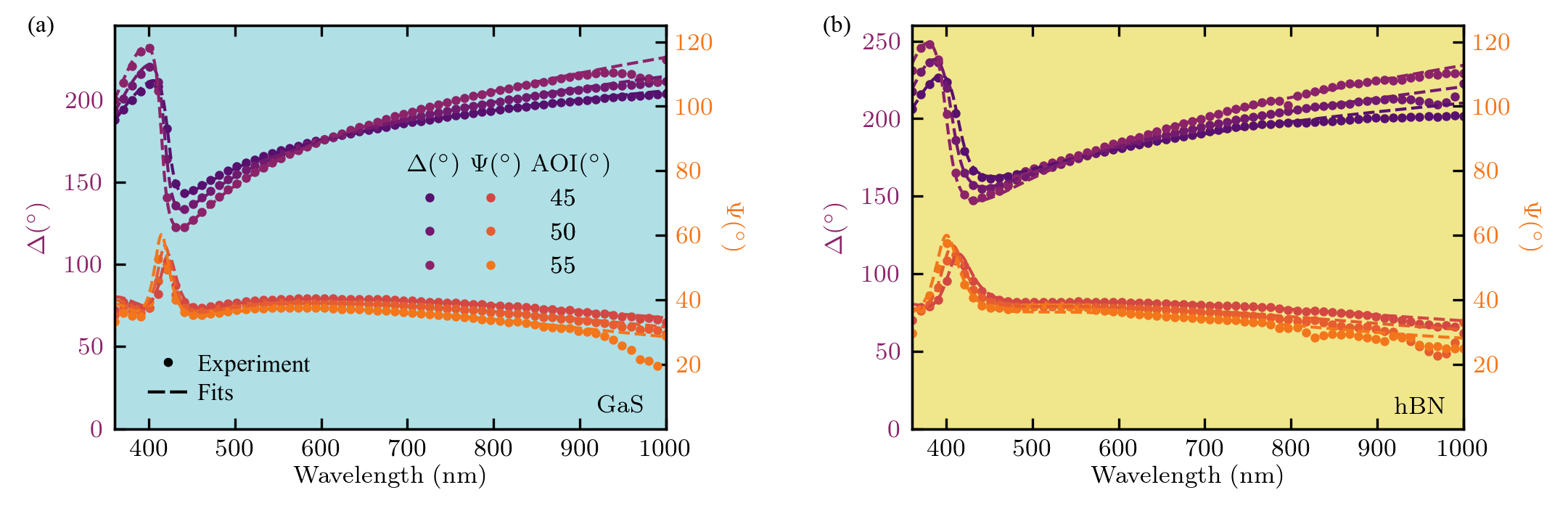}
  \caption{\textbf{Measurement and fits of $\Delta$ and $\Psi$ from the complex reflectance ratio for large bandgap materials.} 
	The $\Delta$ (dark color circles) and $\Psi$ (bright color circles) values are measured at three different incidence angles in a micro-ellipsometry setup and are later fit with a Sellmeier dispersion law (dark and bright color dashed curves respectively) for \textbf{(a)} GaS and \textbf{(b)} hBN.}
\label{Ell1}
\end{figure}

\FloatBarrier

\textbf{Transition metal dichalcogenides} The second group of materials includes all TMDs with bandgaps within the experimental range and also exhibit anisotropic crystal symmetries. In the past, Lorentzian oscillator-based models have been used to approximate the strong excitonic behavior of TMDs in-plane \cite{Verre2019} yet this model lacks the ability to describe the observed spectral asymmetry in absorption and photoluminescence studies. Similar to an earlier report \cite{Ermolaev2021}, we utilize a Tauc-Lorentz oscillator model to describe the exciton resonances ($\varepsilon_{TL,i}$, where i = 1, 2, 3... accounts for multiple excitonic resonances) together with a constant $\varepsilon_{\infty}$ as well as a UV pole ($\varepsilon_{UV}$) term to take into account higher energy electronic transitions. The definition of the in-plane dielectric constant as a function of energy can be written in the following form:

\begin{equation}
\varepsilon_{IP}(E) = \varepsilon_{\infty} + \varepsilon_{UV}(E) + \sum_{i}\varepsilon_{TL,i}(E), \\
\label{eq:2}
\end{equation}

\noindent
where:

\begin{equation}
\begin{array}{lr}
\varepsilon_{UV}(E) = \frac{A_p}{E_{p}^{2}-E^2}, \\
\varepsilon_{TL,i}(E) = \varepsilon_{TL1,i}(E) + i\varepsilon_{TL2,i}(E),
\end{array}
\label{eq:3}
\end{equation}

\noindent
where $E_p$ and $A_p$, extracted from the fitting of the experimental data, represent the position and broadening of the high energy electronic transitions while $\varepsilon_{TL1,i}$ and $\varepsilon_{TL2,i}$ are the real and imaginary parts of the dielectric constant contribution from the multiple excitonic resonances describing refraction and extinction respectively. The imaginary dielectric constant contribution from the excitonic resonances can be further described as follows:

\begin{equation}
\varepsilon_{TL2,i}(E) = 
\left\{    
\begin{array}{lr}
0, \quad\text{for}\quad  E < E_{g,i} \\
\frac{(E-E_{g,i})^2}{E^2} + \frac{A_{i}E_{0}\Gamma_{i}E}{(E_{0,i}^{2}-E^2)^{2}-\Gamma_{i}^{2}E}, \quad\text{for}\quad E \geq E_{g,i}
\end{array}
\right\},
\label{eq:4}
\end{equation}

\noindent
where $A_i$ is the individual exciton oscillator strength, $\Gamma_i$ is the linewidth or broadening and $E_0$ is its energy. The real dielectric constant contribution ($\varepsilon_{TL1,i}(E)$) is derived from the imaginary contribution through Kramers-Kronig integration. As the strong excitonic binding energy in these materials confines these excitations to a single layer, all absorption processes can be ascribed to the in-plane regime. Therefore, in order to describe the out-of-plane contribution to the dielectric constant we use a Cauchy approximation to represent its transparent behavior \cite{Ermolaev2021}:

\begin{equation}
\begin{array}{lr}
n_{OP}^{2}(\lambda) = A + \frac{B}{\lambda^2} + \frac{C}{\lambda^4},\\
k_{OP}^{2}(\lambda) = 0,
\end{array}
\label{eq:5}
\end{equation}

\noindent
where $A$, $B$ and $C$ are fitting parameters. The ellipsometric data fit with the Tauc-Lorentz/UV pole model as well as the Cauchy approximation are displayed in Figure \ref{Ell2}.

\begin{figure}[ht!]
	\centering
  \includegraphics[width=\linewidth]{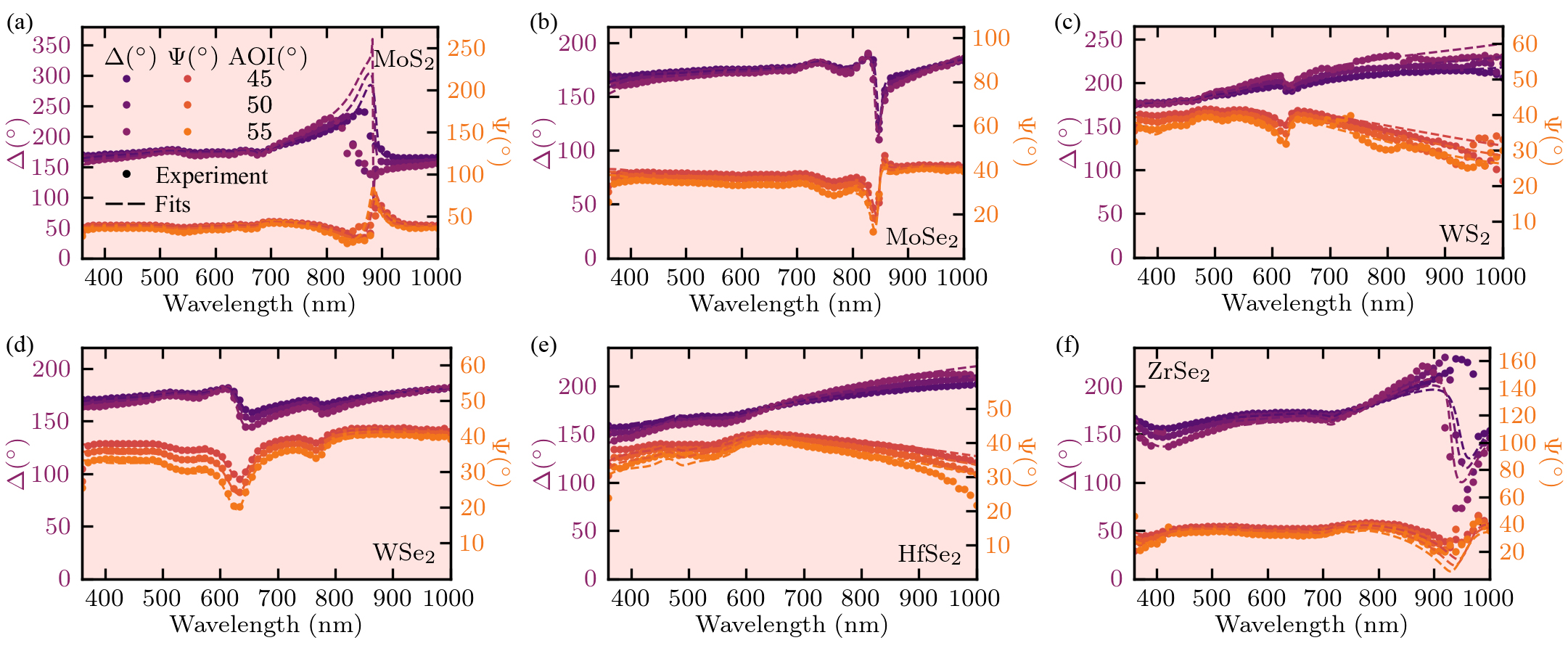}
  \caption{\textbf{Measurement and fits of $\Delta$ and $\Psi$ from the complex reflectance ratio for transition metal dichalcogenides.} 
	The $\Delta$ (dark color circles) and $\Psi$ (bright color circles) values are measured at three different incidence angles in a micro-ellipsometry setup and are later fit with a Tauc-Lorentz/UV pole model (dark and bright color dashed curves respectively) for \textbf{(a)} MoS$_2$, \textbf{(b)} MoSe$_2$, \textbf{(c)} WS$_2$, \textbf{(d)} WSe$_2$, \textbf{(e)} HfSe$_2$ and \textbf{(f)} ZrSe$_2$.}
\label{Ell2}
\end{figure}

\FloatBarrier

\textbf{Isotropic materials} The last group of materials are layered but despite the inherent anisotropy in their crystal structure, their dielectric response is best described by an isotropic model. Due to expected strong (in the case of MnPSe$_3$ and NiPS$_3$) or weak excitonic resonances (in the case of In$_2$Se$_3$ we utilize the Tauc-Lorentz model which we established for the in-plane dielectric response of TMDs in the previous group of materials. The ellipsometry data and model fits are shown in Figure \ref{Ell3}.

\begin{figure}[hb!]
	\centering
  \includegraphics[width=\linewidth]{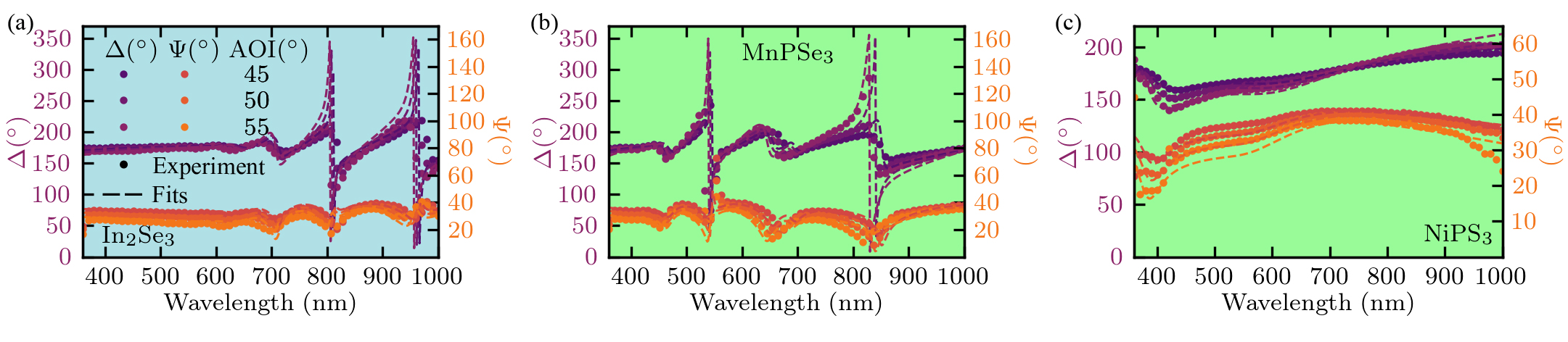}
  \caption{\textbf{Measurement and fits of $\Delta$ and $\Psi$ from the complex reflectance ratio for isotropic materials.} 
	The $\Delta$ (dark color circles) and $\Psi$ (bright color circles) values are measured at three different incidence angles in a micro-ellipsometry setup and are later fit with an isotropic Tauc-Lorentz model (dark and bright color dashed curves respectively) for \textbf{(a)} In$_2$Se$_3$, \textbf{(b)} MnPSe$_3$ and \textbf{(c)} NiPS$_3$.}
\label{Ell3}
\end{figure}

\FloatBarrier

\begin{center}
\Large
\textbf{Supplementary Note 2: Birefringence in van der Waals materials with anisotropic refractive indices}
\end{center}

\begin{figure}[hb!]
	\centering
  \includegraphics[width=\linewidth]{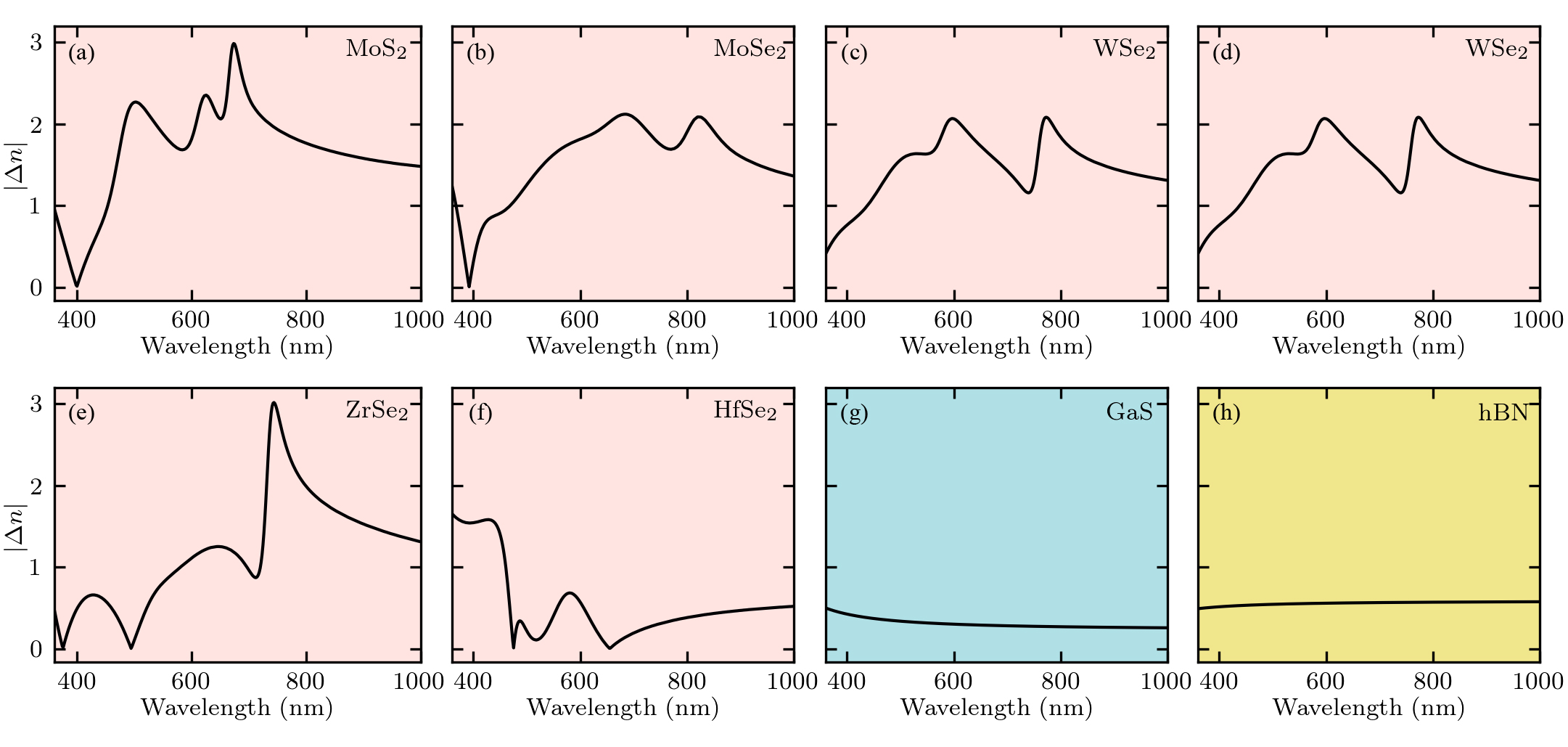}
  \caption{\textbf{Anisotropic material birefringence.}
	Birefringence values ($|\Delta n|$) obtained from the refractive indices ($n$) of materials with an anisotropic dielectric response including \textbf{(a)} MoS$_2$, \textbf{(b)} MoSe$_2$, \textbf{(c)} WS$_2$, \textbf{(d)} WSe$_2$, \textbf{(e)} ZrSe$_2$, \textbf{(f)} HfSe$_2$, \textbf{(g)} GaS, \textbf{(h)} hBN.}
\label{SFigDelN}
\end{figure}

Many of the materials for which we have extracted complex refractive indices exhibit anisotropies in their dielectric response when comparing in-plane and out-of-plane contributions as a result of their crystal symmetry. We have extracted the birefringence ($|\Delta n|$) of each material by subtracting the refractive index of the out-of-plane contribution from that of the in-plane contribution as shown in Figure \ref{SFigDelN}(a)-(h). Particularly noteworthy are the large birefringence values ($\approx$ 3.0) obtained for MoS$_2$ and ZrSe$_2$ near their lowest energy excitonic resonance in the near-infrared range.

\FloatBarrier

\begin{center}
\Large
\textbf{Supplementary Note 3: Fabricated nanostructures in other materials}
\end{center}

\begin{figure}[ht!]
	\centering
  \includegraphics[width=\linewidth]{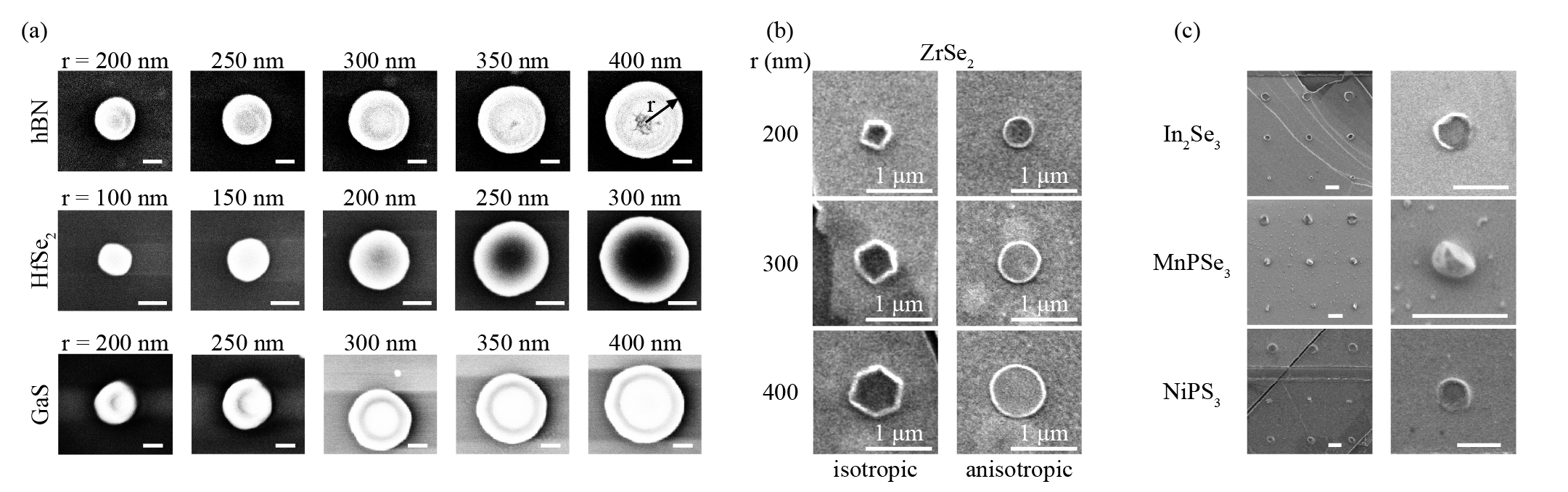}
  \caption{\textbf{Resulting structures after fabrication of a variety of materials.}
	\textbf{(a)} SEM images of fabricated nanopillars with a range of radii in hBN, HfSe$_2$ and GaS after respective isotropic etching recipes yielding a circular geometry. Scale bars = 200 nm.
	\textbf{(b)} SEM images of fabricated nanopillars with a range of radii in ZrSe$_2$ using an isotropic and anisotropic etching recipe yielding hexagonal and circular geometries respectively. A ring-like feature is observed resulting from a different etch rate of the native oxide.
	\textbf{(c)} SEM images of attempted fabrications of nanopillars in In$_2$Se$_3$, MnPSe$_3$ and NiPS$_3$ which resulted in shallow etching and therefore poor nanoantenna formation. Scale bars = 1 $\mu$m.}
\label{SFigFabs}
\end{figure}

In addition to the nanopillars fabricated in the TMDs discussed in the main text, we also attempted to fabricated nanopillars in hBN, HfSe$_2$, ZrSe$_2$, GaS, In$_2$Se$_3$, MnPSe$_3$ and NiPS$_3$. Circular nanopillar structures were fabricated in hBN and HfSe$_2$ using the same isotropic etching recipe as the one described for the fabrication of other TMD nanostructures as shown in the upper two rows of Figure \ref{SFigFabs}(a). For both materials, the fabrication procedure yielded no hexagonal geometries yet the achieved circular geometry was reliably reproducible.

The lowest row in Figure \ref{SFigFabs}(a) shows the results of fabricating GaS nanopillars which required a different etching recipe involving a chlorine gas. For this fabrication, the EBL patterning and development steps were identical as before, however, 5 sccm of SiCl$_4$ gas were used in the etching step with an increased DC bias and reduced chamber pressure. This resulted in the circular geometry shown in the lowest row of Figure \ref{SFigFabs}(a). 

The fabrication of ZrSe$_2$, unlike HfSe$_2$ and MoSe$_2$, yielded both circular as well as hexagonal geometries using the anisotropic and isotropic etch respectively as shown in Figure \ref{SFigFabs}(b), suggesting a crystal axis selectivity of the process. After etching, however, both geometries exhibited a small ring-like feature at the top of each nanopillar which we believe to be a result of a native oxide, known to form in ZrSe$_2$ \cite{Mleczko2017}, that may require different etching conditions in order to be fully removed. This oxide layer is also expected to form on HfSe$_2$ as well, however, it did not maintain a ring-like shape as seen from the middle row of Figure \ref{SFigFabs}(a).

Lastly, we also attempted to etch the rest of the materials for which we extracted a refractive index, namely, In$_2$Se$_3$, MnPSe$_3$ and NiPS$_3$. This resulted in a very poor etch rate which did not reach completely through the crystals and yielded shallow disk-like features which are shown in the Scanning electron microscopy (SEM) images of Figure \ref{SFigFabs}(c). We attempted to etch these materials using both chlorine and fluorine gases with similar results, however, further attempts, beyond the scope of this work, may yield improved results using different gasses or wet etching. 

\FloatBarrier

\begin{center}
\Large
\textbf{Supplementary Note 4: Twisted WS$_2$ homostructure nanoantenna characterization}
\end{center}

\begin{figure}[hb!]
	\centering
  \includegraphics[width=\linewidth]{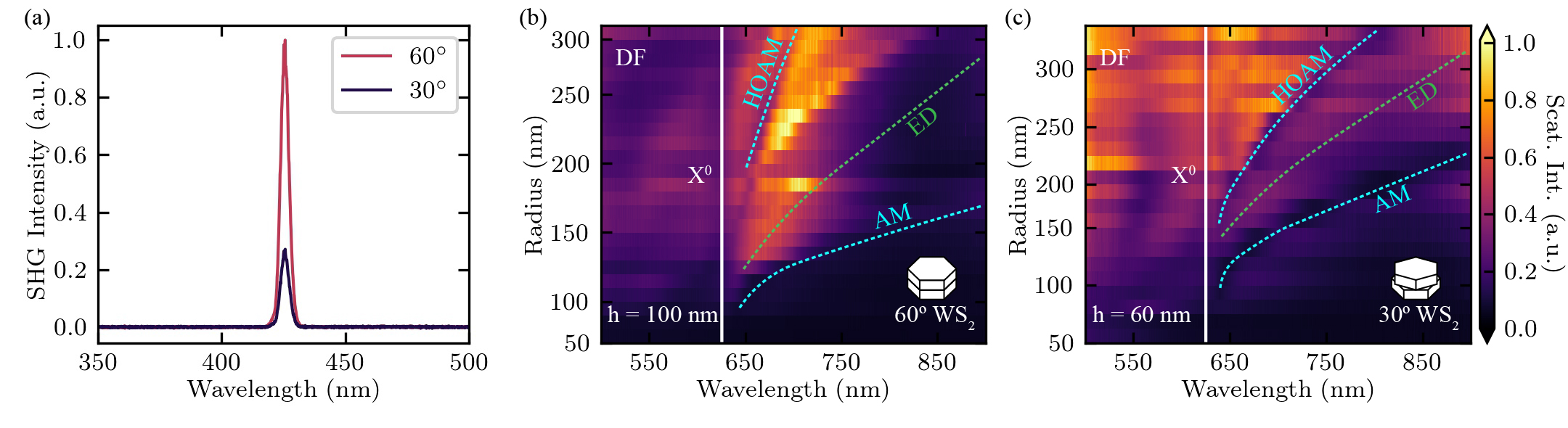}
  \caption{\textbf{SHG and dark field characterization of twisted WS$_2$ homostructure nanoantennas.}
	\textbf{(a)} SHG signal from an unpatterned region of the 60$^\circ$(red) and 30$^\circ$(blue) homostructure under 850 nm femtosecond illumination. 
	\textbf{(b)} Experimental dark field spectra from 60$^\circ$ twisted WS$_2$ monomer nanoantennas (h = 100 nm) yielding an electric dipole (ED), anapole (AM) and higher order anapole mode (HOAM).   
	\textbf{(c)} Experimental dark field spectra from 30$^\circ$ twisted WS$_2$ monomer nanoantennas (h = 60 nm) yielding similar resonances.}
  \label{SI_SHG}
\end{figure}

In order to confirm the difference in twist angles between thin-film homostructures, as shown in Figure 2(c) and (e) of the main text, we employed second harmonic generation (SHG). We excited the two homostructures using a femtosecond Ti-sapphire laser (Mai-Tai) at 850 nm and recorded the SHG signal at 425 nm, shown in Figure \ref{SI_SHG}(a). As the twist angle used in the fabrication of the homostructure locally breaks the centrosymmetry of the WS$_2$ crystal, we expect that the interface between the two constituent thin-film crystals will lead to a dipole allowed second harmonic signal. The SHG intensity is expected to increase with larger twist angles due to the increasingly broken centrosymmetry in neighboring layers at this interface \cite{Yao2021}. The higher second harmonic intensity recorded for the 60$^\circ$ homostructure compared to the 30$^\circ$ sample confirms this.

Next, we recorded the dark field spectra of an array of monomers in each twisted WS$_2$ homostructure. These yielded similar Mie (ED, green), anapole (AM,cyan) and higher order anapole (HOAM, cyan) resonances as expected for WS$_2$ monomer nanoantennas of similar heights as shown in Figure \ref{SI_SHG}(b) and (c) for the 60$^\circ$ and 30$^\circ$ samples respectively. This demonstration shows evidence for the ability to fabricate homo- and heterostructure nanophotonic architectures available only to van der Waals materials. The combination of twist angle and anapole modes in these nanostructures may lead to enhanced SHG signal from the interface \cite{Yao2021,Busschaert2020,Popkova2022,Zotev2022} and therefore insights into interlayer excitons in such thick TMD crystals.  

\FloatBarrier

\begin{center}
\Large
\textbf{Supplementary Note 5: Simulations of monomer and dimer nanoantenna scattering cross sections}
\end{center}

\begin{figure}[ht!]
	\centering
  \includegraphics[width=\linewidth]{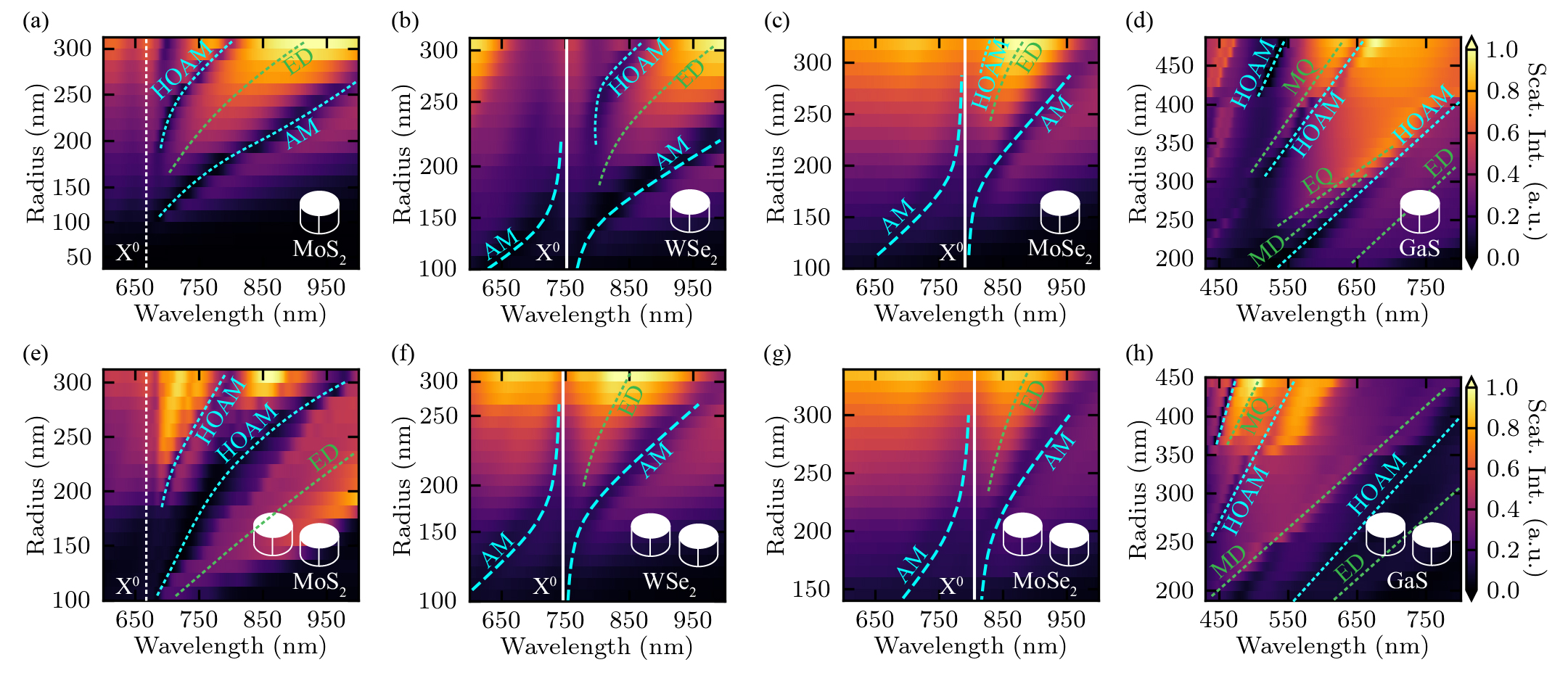}
  \caption{\textbf{Simulations of monomer and dimer nanoantenna scattering cross sections in a variety of 2D materials.}
	\textbf{(a)}-\textbf{(d)} Simulated scattering cross sections for monomer nanoantennas with a range of radii in MoS$_2$ (h = 70 nm), WSe$_2$ (h = 85 nm), MoSe$_2$ (h = 35 nm) and GaS (h = 200 nm). Identified resonances include the neutral exciton resonance of the material (X$^0$, white), the magnetic dipole (MD, green), electric dipole (ED), magnetic quadrupole (MQ) and electric quadrupole (EQ) resonances as well as the anapole (AM, cyan) and higher order anapole mode (HOAM). Observed anticrossings are expected to yield Rabi splittings off 210 meV and 114 meV for the WSe$_2$ and MoSe$_2$ monomers respectively. Similarly the expected Rabi splittings for dimers are 161 meV and 139 meV in WSe$_2$ and MoSe$_2$ respectively.
	\textbf{(e)}-\textbf{(h)} Simulated scattering cross sections for dimer nanoantennas with a range of radii in MoS$_2$ (h = 190 nm), WSe$_2$ (h = 45 nm), MoSe$_2$ (h = 30 nm) and GaS (h = 145 nm). Dimer gaps range from 50 to 200 nm. Observed anticrossings are expected to yield Rabi splittings off 210 meV and 114 meV for the WSe$_2$ and MoSe$_2$ monomers respectively. Identified resonances are similar to those in monomer nanoantennas.}
  \label{SFigDarkFieldSims}
\end{figure}

We have simulated the scattering cross sections expected from the fabricated monomer and dimer structures discussed in the main text, using the finite-difference time-domain (FDTD) technique. The results of these calculations are displayed in Figure \ref{SFigDarkFieldSims}. From the comparison of these simulations to the experimental data we can see good agreement as well as identify Mie resonances, such as the magnetic dipole (MD, shown as a local maximum), electric dipole (ED), magnetic quadrupole (MQ) and electric quadrupole (EQ) resonance as well as the light confining anapole resonances (AM and HOAM, shown as local minima). As expected from Mie theory, all resonances redshift with increasing nanostructure radius. 

For monomer and dimer structures in TMD materials, such as WSe$_2$ and MoSe$_2$, the FDTD simulations also reproduce the anticrossings observed in the experimental spectra. In order to confirm that these anticrossings provide evidence of strong coupling, the measured energy splittings must satisfy the following condition \cite{Dovzhenko2018,Shen2022}:

\begin{equation}
\Omega_{R} > \frac{\left(\frac{1}{\gamma_{X^0}} + \frac{1}{\gamma_{AM}}\right)}{2}
\label{eq:6}
\end{equation}

\noindent
where $\Omega_R$ is the recorded Rabi splitting, $\gamma_{X^0}$ is the lifetime of the neutral exciton and $\gamma_{AM}$ is the lifetime of a photon in the anapole mode. The inverse of the lifetimes of the neutral exciton in each material and the photon in the anapole mode can be extracted from the linewidths (FWHM) of each resonance in the uncoupled regime if we neglect dephasing effects. The FWHM of each resonance was extracted from lorentzian fits to the dark-field spectra of nanostructures in which the neutral exciton and anapole mode were far detuned. This condition is met for the experimental and simulated antricrossings of WSe$_2$ (210 meV) and MoSe$_2$ (114 meV) monomers, confirming the onset of strong light-matter coupling. These simulations yield higher Rabi splittings than the experimental results which we attribute to fabrication imperfections. For dimer structures in WSe$_2$ and MoSe$_2$ we extract Rabi splittings of 161 meV and 139 meV respectively which similarly imply fabrication imperfections may have limited this value in the measured structures discussed in the main text.

\FloatBarrier

\pagebreak
\begin{center}
\Large
\textbf{Supplementary Note 6: Experimental dark field spectra and simulations of monomer and dimer nanoantenna scattering cross sections for WS$_2$ and hBN}
\end{center}

\begin{figure}[ht!]
	\centering
  \includegraphics[width=\linewidth]{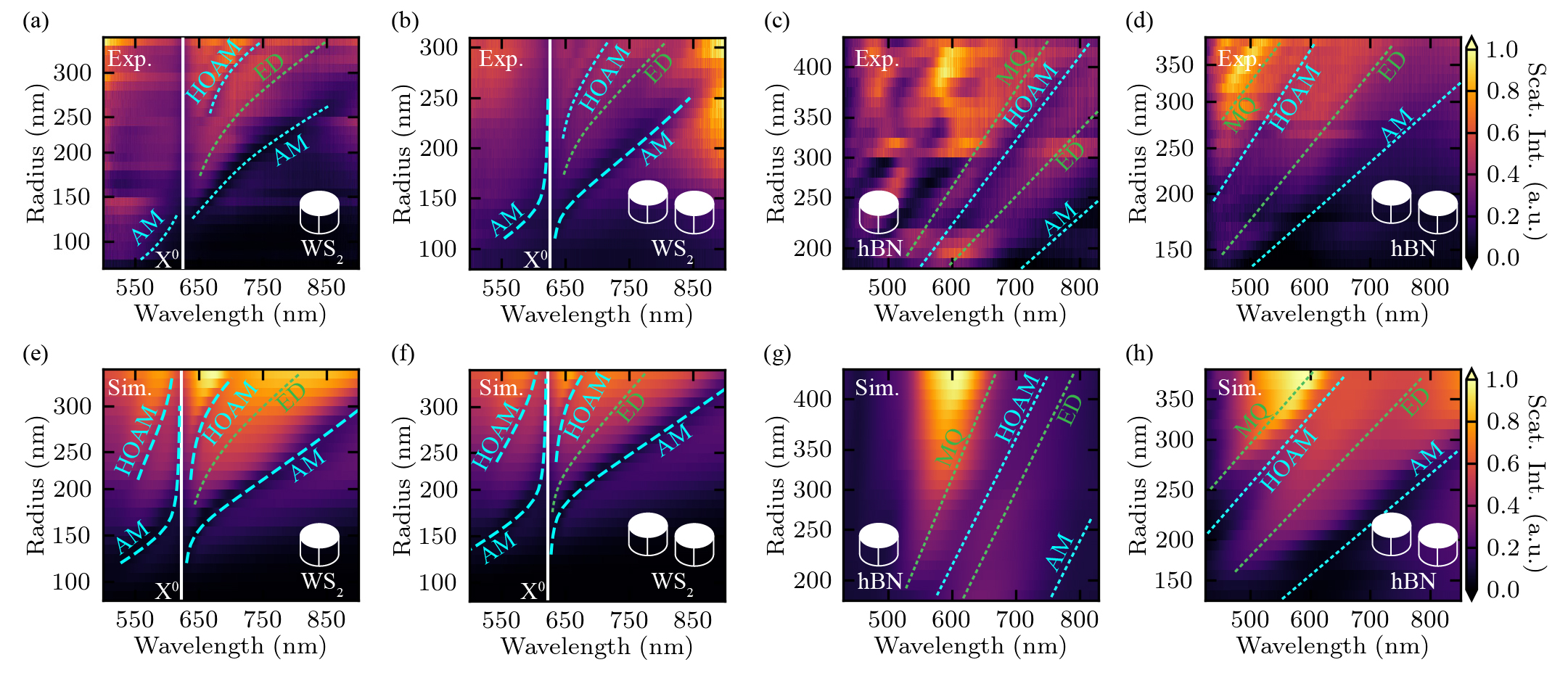}
  \caption{\textbf{Dark field spectra and simulations of monomer and dimer nanoantenna scattering cross sections in WS$_2$ and hBN.}
	\textbf{(a)},\textbf{(b)} Experimental WS$_2$ monomer and dimer nanoantenna (h = 40 nm, g = 50 - 200 nm) dark field spectra. A large anticrossing is observed for the dimer structures confirmed to be intermediate coupling, exhibiting a splitting of 153 meV. 
	\textbf{(c)},\textbf{(d)} Experimental hBN monomer (h = 445 nm) and dimer (h = 200 nm, g = 50 - 200 nm) nanoantenna dark field spectra.
	\textbf{(e)},\textbf{(f)} Simulated WS$_2$ monomer and dimer scattering cross sections. Identified resonances include the neutral exciton resonance of the material (X$^0$, white), the electric dipole resonance (ED, green), the anapole (AM, cyan) and the higher order anapole mode (HOAM). Observed anticrossings are expected to yield Rabi splittings off 177 meV and 179 meV for the monomer and dimer structures respectively.
	\textbf{(g)},\textbf{(h)} Simulated hBN monomer and dimer scattering cross sections. Identified resonances include an electric dipole (ED, green), magnetic quadrupole (MQ) resonance, anapole (AM,cyan) and higher order anapole modes (HOAM).}
  \label{SFigDarkField_WS2_hBN}
\end{figure}

In order to confirm the formation of Mie resonances in more layered materials, we record the dark field spectra and simulate the scattering cross sections for WS$_2$ (h = 40 nm) nanoantennas and hBN monomer (h = 445 nm) and dimer (h = 200 nm) structures as shown in Figure \ref{SFigDarkField_WS2_hBN}. We observe similar Mie and anapole resonances redshifting with increasing radius, as for other materials. We also record small anticrossing near the WS$_2$ neutral exciton for monomers which is confirmed to only exhibit weak coupling due to fabrication imperfections and strong absorption at wavelengths below the neutral exciton resonance. However, we do observe a large anticrossing for the WS$_2$ dimer nanoantennas in Figure \ref{SFigDarkField_WS2_hBN}(b) which yields a Rabi splitting of 153 meV at a radius of 130 nm. This does not meet the condition for strong coupling presented in Supplementary Note 5, however, it does exhibit intermediate coupling as in dimer structures of WSe$_2$ and MoSe$_2$. Simulations of the monomer and dimer WS$_2$ structures yield Rabi splittings of 177 meV and 179 meV respectively confirming that fabrication imperfections play a dominant role in reducing the energy splitting. Additionally, we can observe an anticrossing for the higher order anapole modes near the energy of the neutral exciton, which is not observed for the fabricated structures. These exhibit even higher Rabi splittings of 210 meV and 205 meV for the monomer and dimer structures respectively. We also record a multitude of higher order peaks and dips in the dark field spectra of hBN monomer nanoantennas which we do not observe in simulations and thus may also be a result of fabrication imperfections or resist residues.

\FloatBarrier

\pagebreak
\begin{center}
\Large
\textbf{Supplementary Note 7: Mode confinement comparison for WS$_2$ nanoantennas on a SiO$_2$ and gold substrate}
\end{center}

In order to understand the confinement of nanoantenna resonances inside and in close proximity to the structure in the case of both a gold and SiO$_2$ substrate, we simulate (using a FDTD technique) electric field intensity profiles for an electric dipole (ED) and anapole mode (AM) in a WS$_2$ monomer nanoantenna on the two surfaces as plotted below in Figure \ref{SFigGoldEFieldComp}. For the anapole mode, shown in Figure \ref{SFigGoldEFieldComp}(a) and (c) for a SiO$_2$ and gold substrate respectively, we observe a larger confinement of the resonance inside the nanoantenna in the case of a metallic surface as compared to the low-index dielectric. The maximum electric field intensity in the case of a gold substrate is an order of magnitude higher than for the SiO$_2$ surface. The blueshift of the resonance on the metallic as compared to the low-index dielectric substrate observe in Figure 4 of the main text is attributed to this higher confinement within the nanoantenna volume. 

Similarly for the electric dipole mode, plotted in Figure \ref{SFigGoldEFieldComp}(b) and (d) for a SiO$_2$ and gold substrate respectively, the maximum electric field intensity is an order of magnitude higher in the WS$_2$ monomer on a gold surface, however, the mode confinement is reduced. For a SiO$_2$ substrate, the electric dipole mode is largely confined to the outside edges of the structure as shown in Figure \ref{SFigGoldEFieldComp}(b). In the case of a gold substrate, the ED resonance is shifted towards the nanoantenna-metal interface increasing the mode volume within the nanoantenna, as observed in Figure \ref{SFigGoldEFieldComp}(d). The observed redshift of this resonance for a metallic substrate in Figure 4 of the main text is attributed to this reduced confinement. The overall increased electric field intensity observed for the electric dipole mode in the WS$_2$ nanoantenna on a gold surface, however, suggests additional contributions which may originate from a hybrid Mie-plasmonic resonance \cite{Yang2017}.

\begin{figure}[ht!]
	\centering
  \includegraphics[width=\linewidth]{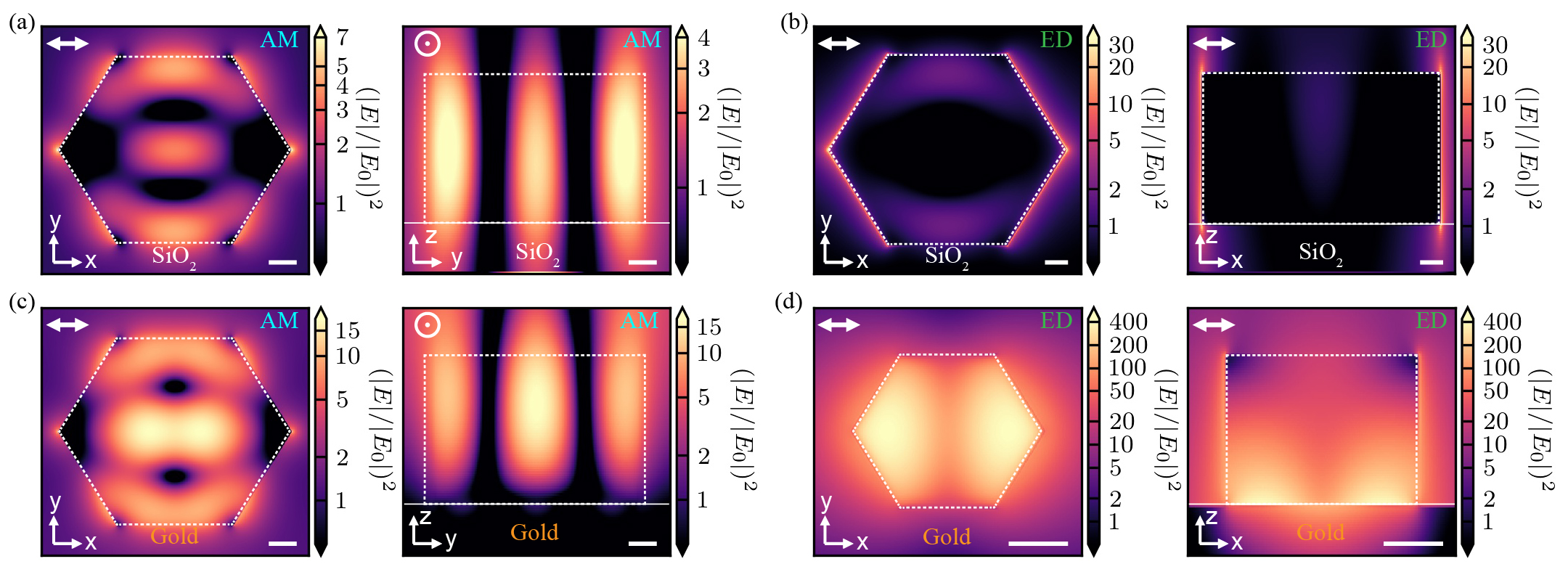}
  \caption{\textbf{Electric field profiles of anapole and electric dipole modes in WS$_2$ nanoantennas on a SiO$_2$ and gold substrate.}
	\textbf{(a)} Horizontal (left panel) and vertical (right panel) cross sectional spatial distribution of the electric field intensity through the middle of a WS$_2$ monomer nanoantenna on SiO$_2$ substrate at the anapole resonance. Mode volume is mostly confined to the inside of the nanoantenna volume. 
	\textbf{(b)} Horizontal (left panel) and vertical (right panel) cross sectional spatial distribution of the electric field intensity through the bottom (left panel) and middle (right panel) of a WS$_2$ monomer nanoantenna on SiO$_2$ substrate at the electric dipole resonance. Mode volume is mostly confined to the outside edges of the nanoantenna.
	\textbf{(c)} Same as \textbf{(a)} for a WS$_2$ monomer nanoantenna on a gold substrate. Mode volume is mostly confined to the inside of the nanoantenna volume.
	\textbf{(d)} Same as \textbf{(b)} for a WS$_2$ monomer nanoantenna on a gold substrate. Mode volume is mostly confined to the nanoantenna-gold interface. Polarization orientation of incident illumination is indicated in the top left corner of each panel. Dashed white outlines represent the physical edges of the structures. Scale bars = 50 nm.}
  \label{SFigGoldEFieldComp}
\end{figure}

\FloatBarrier

\pagebreak
\begin{center}
\Large
\textbf{Supplementary Note 8: Transferable photonics}
\end{center}

\begin{figure}[ht!]
	\centering
  \includegraphics[width=\linewidth]{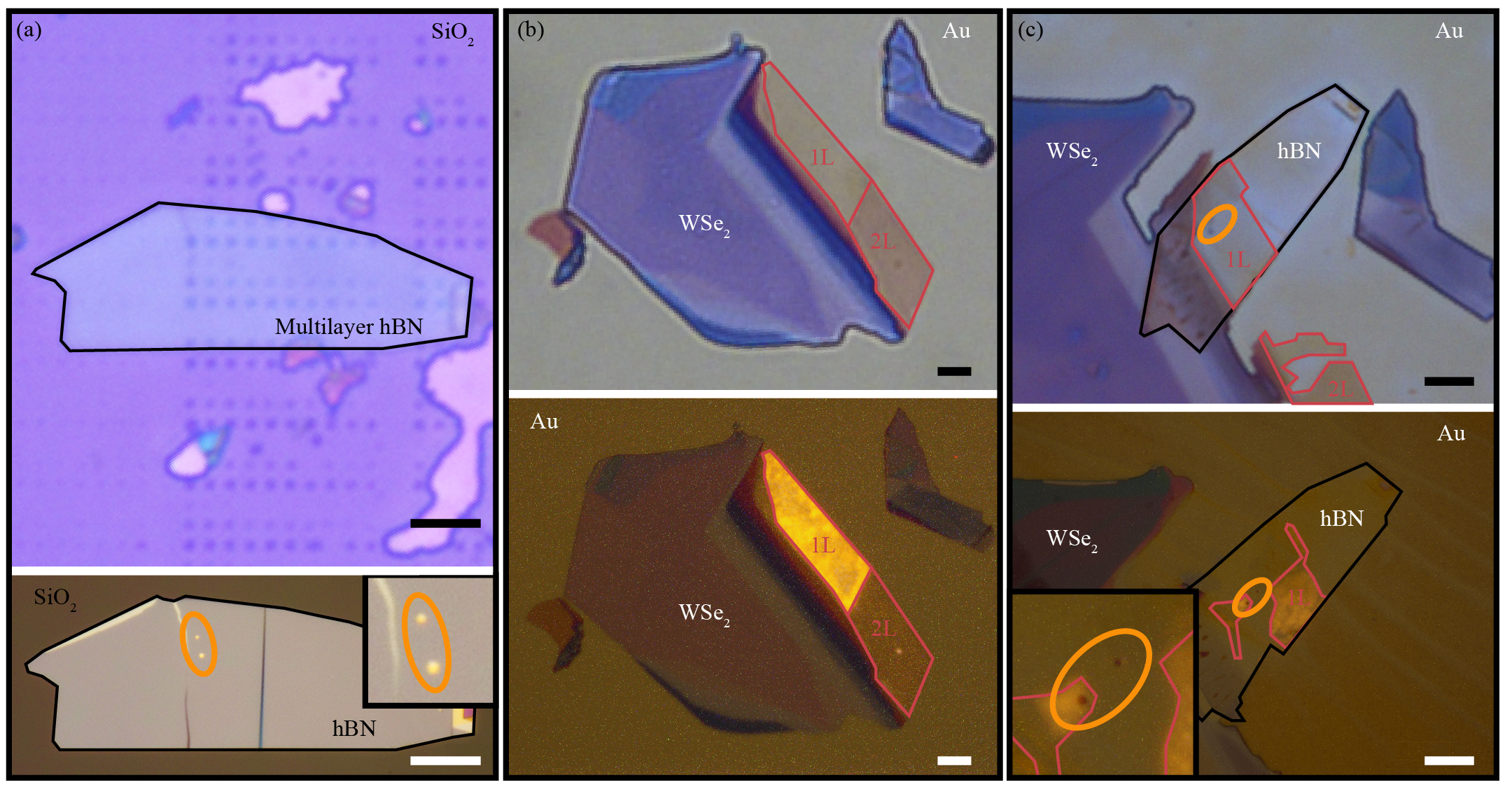}
  \caption{\textbf{WS$_2$ monomer nanoantenna transfer onto a WSe$_2$ monolayer and bilayer on a gold substrate.}
	\textbf{(a)} Upper panel: Bright field microscope image of multilayer hBN exfoliated onto a PMMA/PDMS membrane approaching an array of WS$_2$ monomer nanoantennas on a SiO$_2$ substrate. Lower panel: Microscope image of the same hBN crystal after detaching two nanoantennas from the SiO$_2$ substrate. Yellow circle highlights position of detached nanoantennas. Inset: magnified image at the position of the nanoantennas.
	\textbf{(b)} Bright field (upper panel) and PL image (lower panel) of a transferred monolayer and bilayer WSe$_2$ crystal onto a gold substrate.  
	\textbf{(c)} Bright field (upper panel) and PL image (lower panel) of transferred WS$_2$ nanoantennas with multilayer hBN onto a monolayer WSe$_2$ crystal on a gold substrate. Yellow circle highlights position of detached nanoantennas. Inset: magnified image at the position of the nanoantennas. Scale bars = 10 $\mu$m.}
  \label{SI_TransPhot}
\end{figure}

Due to the weak van der Waals adhesion of the nanoantennas on a SiO$_2$ substrate, we can use an all dry method to pick them up and transfer them onto another substrate which may be sensitive to damage from standard nanofabrication techniques. In order to achieve this, we first exfoliate multilayer hBN onto a PMMA/PDMS membrane placed on a glass slide. Using a micrometer stage transfer setup, we attach the multilayered hBN crystal to fabricated WS$_2$ monomer nanoantennas on a SiO$_2$ substrate, as shown in the upper panel of Figure \ref{SI_TransPhot}(a), at 80$^\circ$C and cool to 30$^\circ$C. After this we slowly peel off the hBN, detaching a few nanoantennas from the SiO$_2$ substrate as shown in the lower panel of Figure \ref{SI_TransPhot}(a). Next, we exfoliate a monolayer and bilayer WSe$_2$ crystal onto a PDMS stamp and slowly approach and attach this to a gold substrate at room temperature. After a slow peel of the PDMS, the WSe$_2$ crystal is transferred onto the gold substrate as shown in Figure \ref{SI_TransPhot}(b). Lastly, we bring the hBN crystal with the attached monomer nanoantennas into contact with the WSe$_2$ monolayer and slowly peel back the PDMS membrane applying a few drops of acetone. This results in a deposition of the PMMA membrane with the hBN and WS$_2$ nanoantennas onto the WSe$_2$ monolayer as shown in Figure \ref{SI_TransPhot}(c). A magnified image of the transferred nanoantennas onto the WSe$_2$ monolayer is shown in the inset of the lower panel of Figure \ref{SI_TransPhot}(c). As some of the acetone has reached the monolayer, the optical quality of the WSe$_2$ single crystal is affected as observed in comparison between the lower panels of Figure \ref{SI_TransPhot}(b) and (c). This technique may also lead to a layer of PDMS contamination between the WS$_2$ nanoantennas and WSe$_2$ monolayer and is not well controlled. Further development is required in order to transfer nanophotonic structures from one substrate onto another in a reliable and scalable way, yet this is beyond the scope of this work. 
 
\FloatBarrier

\pagebreak
\begin{center}
\Large
\textbf{Supplementary Note 9: Third harmonic susceptibility extraction}
\end{center}

We extract the $\chi^{(3)}$ values plotted in Figure 5 of the main text by comparing the measured normalized THG intensity with simulated THG intensity from the following expression:

\begin{equation}
    \chi^{(3)}= \sqrt{\frac{I_{3\omega,exp}/I_{\omega,exp}^3}{I_{3\omega,model}}}, \\
    \label{eq:7}
\end{equation}

\noindent
where $\omega$ and $3\omega$ are the frequency of excitation and third harmonic generation respectively. $I_{\omega,exp}$ and $I_{3\omega,exp}$  are the experimental excitation and THG intensity respectively, while $I_{3\omega,model}$ is the simulated third harmonic intensity which can be written as:

\begin{equation}
I_{THG,model} = \frac{81\omega^{2}}{16\epsilon_{0}^{4}c^{4}n(3\omega)n(\omega)^{3}}\left|\int_{V_{crystal}}\mathbf{\chi}^{(3)}(\omega)\vdots\mathbf{P}_{THG}(\omega)\cdot\mathbf{E_{det}}(3\omega)dV\right|^{2}, \\
\label{eq:8}
\end{equation}

\noindent
where $\epsilon_0$ and $c$ are the permittivity and speed of light in vacuum respectively. The prefactors of equation \ref{eq:8} originate due to a plane wave approximation, via an one dimensional Green’s function for the Helmholtz equation, to model a dipole moment at the detector in nonlinear scattering theory \cite{OBrien2015}. Using nonlinear scattering theory \cite{OBrien2015}, we can also define the nonlinear polarizability ($\mathbf{P}_{THG}(\mathbf{r},\omega)$) of the THG process leading to emission at a frequency of $3\omega$ from each thin-film material as follows:

\begin{equation}
\mathbf{P}_{THG}(\mathbf{r},\omega) = \epsilon_0 \mathbf{\chi}^{(3)}(\omega)\mathbf{E^3_{exc}}(\mathbf{r},\omega). \\
\label{eq:9} 
\end{equation}

This model requires knowledge of the linear optical properties of each material at the excitation ($n(\omega)$) and THG frequency ($n(3\omega)$) as well as the thickness of the illuminated crystal ($V_{crystal}$) recorded via atomic force microscopy. The values of the expected THG intensity were calculated using the above nonlinear scattering theory \cite{OBrien2015} in conjunction with two linear transfer-matrix method simulations.

\bibliographystyle{unsrt}
\bibliography{./library}